\documentclass[11pt]{article}
\usepackage{graphicx} 
\usepackage{caption}
\usepackage{subcaption}
\usepackage{amsmath}
\usepackage[T1]{fontenc}
\usepackage[utf8]{inputenc}
\usepackage{authblk}
\usepackage{hyperref}
\usepackage{multirow}
\DeclareMathOperator*{\argmax}{argmax} 
\DeclareMathOperator*{\argmin}{argmin} 
\usepackage{enumitem}
\usepackage{mathtools}

\DeclarePairedDelimiter\floor{\lfloor}{\rfloor}

\title{Unraveling S$\&$P500 stock volatility and networks - An encoding-and-decoding approach}
\author[1]{Xiaodong Wang}
\author[1]{Fushing Hsieh}
\affil[1]{Department of Statistics, University of California, Davis.}
\date{} 

\begin{document}

\maketitle

\section*{Abstract}

Volatility of financial stock is referring to the degree of uncertainty or risk embedded within a stock’s dynamics. Such risk has been received huge amounts of attention from diverse financial researchers. By following the concept of regime-switching model, we proposed a non-parametric approach, named encoding-and-decoding, to discover multiple volatility states embedded within a discrete time series of stock returns. The encoding is performed across the entire span of temporal time points for relatively extreme events with respect to a chosen quantile-based threshold. As such the return time series is transformed into Bernoulli-variable processes. In the decoding phase, we computationally seek for locations of change points via estimations based on a new searching algorithm in conjunction with the information criterion applied on the observed collection of recurrence times upon the binary process. Besides the independence required for building the Geometric distributional likelihood function, the proposed approach can functionally partition the entire return time series into a collection of homogeneous segments without any assumptions of dynamic structure and underlying distributions. In the numerical experiments, our approach is found favorably compared with parametric models like Hidden Markov Model. In the real data applications, we introduce the application of our approach in forecasting stock returns. Finally, volatility dynamic of every single stock of S\&P500 is revealed, and a stock network is consequently established to represent dependency relations derived through concurrent volatility states among S\&P500.

\section{Introduction}
To discover the mystery of the stock dynamics, financial researchers focus on stock returns or log returns. Black and Scholes \cite{1} proposed in their seminal work to use stochastic processes in modeling stock prices. One particular model of focus is the geometric Brownian motion (GBM), which assumes all log-returns being normally distributed. That is, if a time series of the price of a stock is denoted as $\{X(t)\}_t$, the GBM modeling structure prescribes that

$$log\frac{X(t)}{X(t-1)} \sim N(\mu,\sigma^2)$$

Later Merton \cite{2} extended Black and Scholes’ model by involving time-dependent parameters for accommodating potential serial correlations. Further, in order to go beyond normal distribution, models belonging to a broader category, including a general L\'evy process or particular geometric L\'evy process model \cite{11}, become popular and appropriate alternatives by embracing stable distribution with heavy tails. Since the independent increments property of Brownian motion or L\'evy process, returns over disjoint equal-length time intervals remain identically independently distributed (i.i.d). So, the independent increment property restricts modeling stochasticity to be invariant across the entire time span. However, it is well known that the distributions of returns are completely different over various volatility stages. Thus, these models are prone to fail in capturing extreme price movements \cite{3}.

Research attempts from various perspectives have experimented to make stock price modelings more realistic. One fruitful perspective is to incorporate stochastic volatility into stock price modeling. From this perspective, regime-switching or hidden-state models are proposed to govern the stock price dynamics. The regime-switching model can be represented by the distributional changes between a low-volatility regime and a more unstable high-volatility regime. In particular, different regimes are characterized by distinct sets of distributional modeling structures. One concrete example of such modeling is the Hidden Markov Model(HMM). HMM appeals increasingly to researchers due to its mathematical tractability under the assumption of Markovian. Its parametric approach has gained popularity and its parameter estimation procedures have been discussed comprehensively. For instance, Hamilton \cite{4} described AR and ARCH-type models under Markov regime-switching structure. Hardy \cite{3} offered Markov regime-switching lognormal model by assuming different normal distribution within each state,
$$log\frac{X(t)}{X(t-1)} |s \sim N(\mu_s,\sigma^2_s)$$
where $s$ indicates the hidden states for $s=1,2,...$. Fine et al. \cite{5} developed hierarchical HMM by putting additional sources of dependency in the model. To further increasing the degree of flexibility in modeling stochastic volatility, another well-known financial model was related to volatility clustering \cite{19}. For instance, GARCH models have been studied to model the time-varying conditional variance of asset returns \cite{20}. However, such a complicated dynamic structure usually involves a large number of parameters. This modeling complexity renders the model hard to interpret. In contrast, non-parametric approaches are still scarce in the literature due to the lack of tractability and involvement of many unspecified characteristics \cite{12}.

In this paper, we take up a standpoint right in between the purely parametric and non-parametric modelings. We adopt the research platform of regime-switching models but aim to develop an efficient non-parametric procedure to discover the dynamic volatility without assuming any distribution family nor underlying structure. The idea is motivated by a non-parametric approach, named Hierarchical Factor Segmentation(HFS) \cite{7,8}, to mark extremely large returns as 1 and others 0, and then partition the resultant 0-1 Bernoulli sequence into alternating homogeneous segments. HFS takes advantage in transforming the returns into a 0-1 process with time-varying Bernoulli parameters, so parametric approaches such as likelihood-based function can be applied to fit each segment respectively. However, it is unclear in HFS to define a ``large'' return that should be marked, which makes the implementation limited in application. Another limitation of HFS, which is also shared by regime-switching models or HMM, is that there exists no data-driven way of determining the number of underlying regimes or hidden states.

We propose an encoding-and-decoding approach to resolve the issues tied to the aforementioned limitations simultaneously. The encoding procedure is done by iteratively marking the returns at different thresholding quantile levels, so the time series can be transformed into multiple 0-1 processes. In the decoding phase, a searching algorithm in conjunction with model selection criteria is developed to discover the dynamic pattern for each 0-1 process separately. Finally, the underlying states are revealed by aggregating the decoding results via cluster analysis. It is remarked that the non-parametric approach is able to discover both light-tail and heavy-tail distributional changes without assuming any dynamic structure or Markovian properties. Though the proposed method is derived under independence or exchangeability conditions, our numerical experiments show that the approach still works for settings with the presence of weak serial dependency, which can be checked by testing the significance of lagged correlation in practice.

Another contribution of this paper is that a searching algorithm is developed to partition a 0-1 process into segments with different probability parameters. Therefore, our computational development is a change point analysis on a sequence of Bernoulli variables with the number of change points being large and unknown. For such a setting and its like, the current searching algorithm is infeasible, such as bisection procedures \cite{21,22}. As an alternative to the hierarchical searching strategy, our proposed search algorithm concurrently generates multiple segments with only a few parameters. The optimal partition of homogeneous segments is ultimately obtained via model selection.

The paper is constructed as follows. In Section2, we review the HFS and develop a new searching algorithm that can handle multiple-states decoding. In Section3, we present the main approach in modeling distributional changes. In Section4, real data analysis is performed to illustrate the procedure of stock forecasting and networks of revealing relational patterns among S$\&$P500. Several remarks and conclusion are given in Section5.

\section{Volatility Dynamics}

To investigate stock dynamics, we consider volatility as a temporal aggregation of rare events which have large absolute returns. Consider the recursive time or waiting time between successive extreme returns, Chang et al.\cite{10} proved that the empirical distribution of recursive time converges to a geometric distribution asymptotically, when the observations are i.i.d, or more generally, in exchangeable join distributions. Under the framework of regime-switching model, each regime can be considered as a homogeneous time period with exchangeable distributions. Motivated by that, geometric distributions with different emission or intensity probabilities are adopted to model returns under different volatility periods. The potentially feasible assumption we made here is the invariance of the recursive time distributions embedded within a regime. Long-term serial correlation is out of our consideration in the paper. Though this popular assumption may seemingly hold when the duration of each regime is short enough, it would raise the switching frequency between different volatility periods. This characteristic makes the model complex and non-traceable. Further, based on the abundant permutation tests in 30-s, 1-min, 5-min return scales, it is claimed in \cite{9} that returns should only be considered exchangeable within a short time period, which should not be longer than, for example, 20 minutes. On the other hand, a longer duration of volatility period would have more samples to ensure a goodness-of-fit of within-regime distributions.

With the above background knowledge in mind, we go into the encoding phase of computational developments. Consider a pair of thresholds $(l, u)$ applied to define events of interest or large returns. Specially, an excursion 0-1 process $C(t)$ at time $t$ is defined as
\begin{equation} \label{eq.thresh}
    C(t) = \begin{cases} 1 & \quad log\frac{X(t)}{X(t-1)} \le l \enspace \textit{or}~log\frac{X(t)}{X(t-1)} \ge u \\ 0 & \quad \textit{otherwise} \end{cases}
\end{equation}
\noindent where $l$ and $u$ are lower and upper quantiles of log returns, respectively. It is easy to generalize the thresholds to involve one-sided tale excursions, for instance, to set $l=0$ and $u \in (0,1)$ to focus on positive returns or upper tail excursions. If the thresholds are set too extreme, then only fewer excursive returns can stand out. As a result, the excursion process is too simple to preserve enough information about the volatility dynamics due to the reduction of sample size. While, if the quantile value is set close to the median, then the dynamic pattern is overwhelmed by irrelevant information or noise. There is an inevitable trade-off between the magnitude of the sample size and the amount of information about excursive events. Our remedy to this problem is to systematically apply a series of thresholds and encode the time series returns into multiple binary (0-1) excursion processes. For the completeness of the analysis, we will discuss a searching algorithm in conjunction with model selection criteria in the section below, which is the key in the decoding phase. More details about the encoding procedure are described later in Section3.

\subsection{The Searching Algorithm}
Suppose a 0-1 excursion process has been obtained. In this subsection, we discuss how to search for potential temporal segmentation. As the study involving multiple change points, we aim to detect abrupt distributional changes from one segment of low-volatility regime to another segment of high-volatility regime. To properly accommodate a potentially large number of unknown change points due to the recursive emissions of volatility, and to effectively differentiate the alternating volatility-switching patterns, the Hierarchical Factor Segmentation(HFS) was employed to partition the excursion process into a sequence of high and low event-intensity segments \cite{8}. 
The original HFS assumes that there exist only two kinds of hidden states within the returns corresponding to low-volatility and high-volatility regimes. Though the assumption is plausible within a short time period, its potential becomes limited when the time series of returns is lengthy and embracing more complicated regime-specific distributions. In this subsection, we expand the HFS by incorporating a more generalized searching algorithm to handle the scenarios of multiple states.

Denote the entire 0-1 excursion process sequence of length $n$ as $\{C(t)\}_{t=1}^n$. The recursive recurrent time between two successive 1's of $\{C(t)\}_{t=1}^n$ is recorded into a sequence, denoted as $\{R(t)\}_{t}$. It is noted that the recurrent time can be 0 if two 1's appear consecutively. Also, we denote $R(1)=0$ if $C(1)=1$. As such, the length of $\{R(t)\}_{t=1}^{n^*}$ is $n^*=n^{'}+1$ where $n^{'}$ is the number of 1's in $\{C(t)\}_{t=1}^n$. To make the notations consistent, we denote $\{C_i(t)\}_{t=1}^{n^*}$ as the $i$-th coding sequence if $i$ is present and its corresponding recurrent time sequence as $\{R_i(t)\}_{t=1}^{n_i^{**}}$.

Suppose that the number of internal states is $m$ and $m >1$. Then, there are $m$ tuning parameters are required in the searching algorithm given below. Denote the first thresholding parameter vector as $T=(T_1,T_2,...,T_{m-1})$ where $T_1<T_2<...<T_{m-1}$, and the second thresholding parameter as $T^{*}$. The searching algorithm is described in Alg.1.

\noindent\rule{12.5cm}{0.8pt}\\
\textbf{Alg.1} multiple-states searching algorithm\\
\rule{12.5cm}{0.4pt}\\
1.Define events of interest and encode the time series of return into a 0-1 digital sequence $\{C(t)\}_{t=1}^{n}$ with 1 indicating an event and 0 otherwise.\\
2.Calculate the recurrence time in $\{C(t)\}_{t=1}^{n}$ and denote the resultant sequence as $\{R(t)\}_{t=1}^{n^*}$.\\
3. For loop: cycle through $i=1,2,...,m-1$:
\begin{enumerate}[label=\roman*.]
    \item Transform $\{R(t)\}_{t=1}^{n^*}$ into a new 0-1 digital strings $\{C_i(t)\}_{t=1}^{n^*}$ via the second-level coding scheme:
\[ C_i(t) = \begin{cases} 1 & \quad R(t)\ge T_i\\
0 & \quad \textit{otherwise} \end{cases}\]

    \item Upon code sequence $\{C_i(t)\}_{t=1}^{n^*}$, take code digit 1 as another new event and recalculate the event recurrence time sequence $\{R_i(t)\}_{t=1}^{n_i^{**}}$

    \item If a recursive time $R_i(t) \ge T^{*}$,
then record its associated time segment in $\{C_i(t)\}_{t=1}^{n^*}$, denoted as $Seg_i$ where $Seg_i \subset \{1,...,n\}$.
\end{enumerate}
4. The $m$ internal states are returned by $S_1=Seg_1$, $S_2=Seg_2\backslash Seg_1$,..., $S_{m-1}=Seg_{m-1}\backslash Seg_{m-2}$, and $S_m=\{1,...,n\}\backslash Seg_{m-1}$.\\
\rule{12.5cm}{0.8pt}\\

\begin{figure}[!h]
\centering
\includegraphics[width=4.5in]{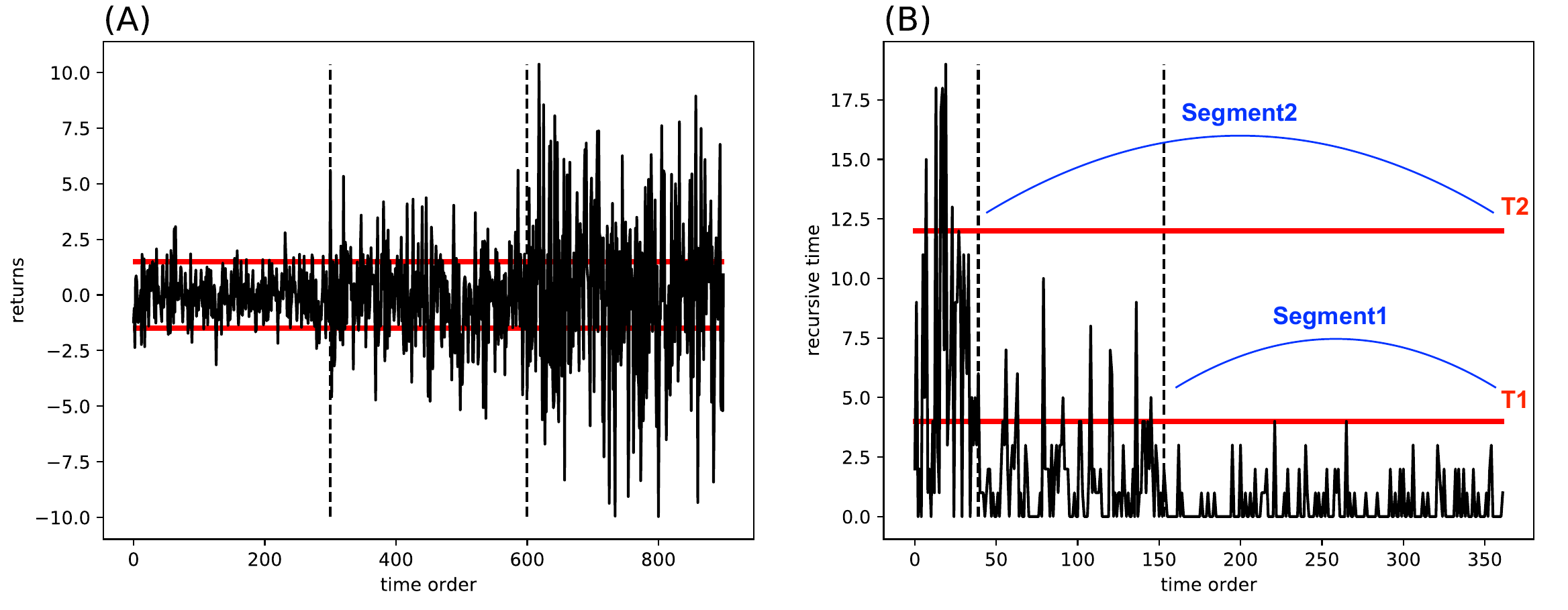}
\caption{A simple example to illustrate the implementation of Alg.1.}
\label{fig.HFS}
\end{figure}

A sequence of Gaussian distributed observations are generated with mean 0 and variance varying under different unknown states in Figure\ref{fig.HFS}(A). A pair of thresholds $l=-2$ and $u=2$ are applied to code the observations via (\ref{eq.thresh}), so a sequence of recursive time is obtained in Figure\ref{fig.HFS}(B). The first-level parameter $T_i$ are set to control the event-intensity that we aim to partition for $i=1,...,m-1$, see thresholds $T_1$ and $T_2$ in Figure\ref{fig.HFS}(B). If $T_i$ takes its maximum $T_{m-1}$, then a high-intensity segment is separated from other level segments, see $T_2$ in Figure\ref{fig.HFS}(B). By decreasing the value of $T_i$ from $T_{m-1}$ to $T_1$ to implement a series of partitions, multiple intensity levels of phases get separated. In this example, $T_1$ is set to partition high- and median-intensity from the low-intensity segment.

In the second level of recursive time calculation, $\{R_i(t)\}_{t=1}^{n_i^{**}}$ are calculated for $i=1,...,m-1$. If $R_i(t)$ is above the second-level threshold $T^{*}$, the segment corresponds to a period with low-intensity events. So, for a fixed $T_i$, $T^{*}$ is set to decide which phases having relatively low intensity, so the rests are in high intensity. It is noticed that $Seg_j \subset Seg_i$, for $j>i$. It is because if a recursive time $R(t)$ is greater than $T_j$, it is greater than $T_i$ as well. By applying the same parameter $T^{*}$ in Figure\ref{fig.HFS}(B), for example, Segment2 is wider than Segment1, so the median-intensity segment is obtained by $Seg2 \backslash Seg1$. More numerical experiment for the application of Alg.1 is available in Appendix\ref{appendix:alg.1}.

\subsection{Mixed-geometric Model}

Multiple volatility phases of time series $S_i$, $i=1,...,m$ are computed and achieved by applying the searching algorithm. Assuming join distribution is exchangeable within each volatility phase, the recursive distribution $\{R(t)\}_{t \in S_i}$ converges to a geometric distribution with parameter $p_{S_i}$ as the sample size going to infinity where $p_{S_i}$ is the emission probability under state $S_i$. Maximized Likelihood Estimation(MLE) and Method of Moment(MOM) give the same estimator for $p_{S_i}$,

\begin{equation} \label{eq.p}
\hat{p}_{S_i}= \frac{1}{\sum_{t \in S_i} R(t)}
\end{equation}
\noindent for $i=1,...,m$. The searching algorithm actually advocates a way to partition the process into segments with $m$ different intensities. With enough sample size, geometric distribution is appropriate to model the $m$ phases with estimated parameter $\hat{p}_{S_i}$. Thus, maximum likelihood approaches can be applied for model selection. A penalized likelihood or loss function can be simplified by,

\begin{equation}
Loss(\theta) = -2 \sum_{i=1}^{m} [\sum_{t \in S_i^{\theta}}{C(t)}log\hat{p}_{S^{\theta}_i} + \sum_{t \in S_i^{\theta}}{(1-C(t))}log(1-\hat{p}_{S^{\theta}_i})] + k N
\end{equation}
\noindent where $S^{\theta}_i$ is generated segments by applying $m$ parameter $\theta=T_1,...,T_{m-1},$ and $T^{*}$; $m$ is the number of embedded phases or hidden states; $N$ is the total number of alternating segments, $N \ge m$, and $k$ is a penalty coefficient. For instance, $k=2$ corresponds to Akaike Information Criterion(AIC), and $k=log(n)$ corresponds to Bayesian Information criterion(BIC). In this paper, we consistently use BIC in all the experiment. The optimal parameters $\theta^*$ are tuned such that the loss can achieves its minimum, so

\begin{equation}
\theta^*=\argmin_{\theta} Loss(\theta)
\end{equation}
Thus, the segments are ultimately achieved by applying $\theta^*$. The computation cost is expensive if all possible $T_1,...,T_{m-1}$ combinations are considered. In practice, a random grid-search strategy can be applied.


\subsection{Simulation: Bernoulli-distributed Observations}

Numerical experiments are conducted to demonstrate the performance of Alg.1. The first experiment is designed to investigate the asymptotic property of the proposed algorithm as sample size $n$ increases. 2 hidden states $S_1$ and $S_2$ are generated in a sequence like $\{S_1,S_1,...,S_1,S_2,...,S_2,S_1,...\}$. The length of each segmentation of a hidden state is a fixed proportion of $n$. The change points are set at $0.1n, 0.2n, 0.4n, 0.7n, 0.9n$, so there are 7 alternative segments in total. Observations are Bernoulli distributed with emission probability $p_1$ under state $S_1$ and $p_2$ under $S_2$. The experiment is repeated via different $n$ and emission $p_1$ and $p_2$.

The mean and standard deviation of decoding error rates for AIC and BIC are presented in Table\ref{tab.asymptotic}. It seems that asymptotic property can hold for the decoding algorithm. Moreover, AIC performs better than BIC in the most cases, especially when $p_1$ is close to $p_2$. We will consistently apply AIC in the rest experiments of the paper.

\begin{table}[h]
\centering
\caption{Independent Bernoulli sequence: average decoding error rates with standard deviation in brackets}
\scalebox{0.7}{%
\begin{tabular}{c|cc|cc|cc|cc|}
\cline{2-9}
                           & \multicolumn{2}{c|}{$p_1$=0.1, $p_2$=0.05} & \multicolumn{2}{c|}{$p_1$=0.1,   $p_2$=0.2} & \multicolumn{2}{c|}{$p_1$=0.1, $p_2$=0.3} & \multicolumn{2}{c|}{$p_1$=0.1, $p_2$=0.5} \\ \hline
\multicolumn{1}{|c|}{n}    & AIC               & BIC              & AIC               & BIC               & AIC              & BIC              & AIC              & BIC              \\ \hline
\multicolumn{1}{|c|}{1000} & 0.361 (0.107)     & 0.455 (0.072)    & 0.280 (0.122)     & 0.410 (0.105)     & 0.115 (0.051)    & 0.147 (0.095)    & 0.060 (0.023)    & 0.054 (0.019)    \\
\multicolumn{1}{|c|}{2000} & 0.308 (0.117)     & 0.441 (0.100)    & 0.183 (0.098)     & 0.309 (0.151)     & 0.073 (0.035)    & 0.059 (0.026)    & 0.035 (0.013)    & 0.030 (0.010)    \\
\multicolumn{1}{|c|}{3000} & 0.235 (0.116)     & 0.412 (0.116)    & 0.122 (0.065)     & 0.190 (0.135)     & 0.048 (0.021)    & 0.042 (0.016)    & 0.023 (0.008)    & 0.020 (0.007)                \\ \hline
\end{tabular}
}
\label{tab.asymptotic}
\end{table}

In the second experiment, data is generated under Hidden Markov Model (HMM) with 2 hidden states. The decoding error rate is calculated and compared for the proposed algorithm and the HMM. Since the true transition probability and emission probability are unknown in the application of Viterbi's\cite{13}, forward-backward algorithm\cite{Baum} and EM algorithm in Baum–Welch type\cite{6} need to be firstly implemented to calibrate the model parameters. Generally, transition and emission are randomly initialized at first and then updated through the iteration. We name the whole process of parameter estimation and decoding by `HMM'. On the other hand, a pure Viterbi's is implemented with the input of true transition and emission, which can be regarded as mathematical `Truth'. 

For the convenience of simulation, we set $p_{11}=p_{22}$ in transition probability matrix $A$, so $A$ is controlled by a only parameter $p_{12}$,
\[ A= \begin{bmatrix}
1-p_{12} & p_{12} \\
p_{12} & 1-p_{12}
\end{bmatrix} \]

Observations of length $n=1000$ are simulated via different transition matrix $A$ and emission $p_1$ and $p_2$. The experiment is repeated and the mean of decoding error rates is reported in Table\ref{tab.Viterbi}. It shows that the overall decoding errors decrease as $p_{12}$ decreases, and the proposed method gets largely improved and close to the `Truth' especially when $p_{12}$ is less than $0.01$. It can be explained by the fact that a simple model with fewer alternating segments is favored for all the approaches. However, Viterbi's in conjunction with the parameter estimation is unsatisfactory in all the cases. It reveals the challenges of applying Viterbi's or Baum–Welch's in reality.

\begin{table}[h]
\centering
\caption{Bernoulli sequence under HMM: average decoding error rates}
\scalebox{0.75}{%
\begin{tabular}{l|ccc|ccc|ccc|ccc|}
\cline{2-13}
\multicolumn{1}{c|}{} & \multicolumn{3}{c|}{$p_1$=0.1, $p_2$=0.05} & \multicolumn{3}{c|}{$p_1$=0.1,   $p_2$=0.2} & \multicolumn{3}{c|}{$p_1$=0.1,   $p_2$=0.3} & \multicolumn{3}{c|}{$p_1$=0.1,   $p_2$=0.5} \\ \hline
\multicolumn{1}{|l|}{$p_{12}$}                     & Truth    & HMM    & Alg.1      & Truth    & HMM    & Alg.1       & Truth    & HMM    & Alg.1       & Truth    & HMM    & Alg.1       \\ \hline
\multicolumn{1}{|l|}{0.1}                     & 0.4558      & 0.4660      & 0.4585   & 0.4411      & 0.4482      & 0.4570    & 0.3572      & 0.3985      & 0.4414    & 0.2162      & 0.2854      & 0.3891    \\
\multicolumn{1}{|l|}{0.05}                    & 0.4225      & 0.4519      & 0.4318   & 0.4092      & 0.4415      & 0.4309    & 0.2945      & 0.3957      & 0.3790    & 0.1430      & 0.2752      & 0.2887    \\
\multicolumn{1}{|l|}{0.01}                    & 0.3590      & 0.4040      & 0.3522   & 0.2839      & 0.4042      & 0.2972    & 0.1431      & 0.3936      & 0.2047    & 0.0415      & 0.2677      & 0.1048    \\
\multicolumn{1}{|l|}{0.005}                   & 0.2777      & 0.3417      & 0.2995   & 0.2086      & 0.3731      & 0.2287    & 0.0639      & 0.3682      & 0.1133    & 0.0168      & 0.2767      & 0.0596    \\ \hline
\end{tabular}
}
\label{tab.Viterbi}
\end{table}

\begin{figure}[h]
    \centering
    \begin{subfigure}[b]{0.45\textwidth}
        \includegraphics[width=\textwidth]{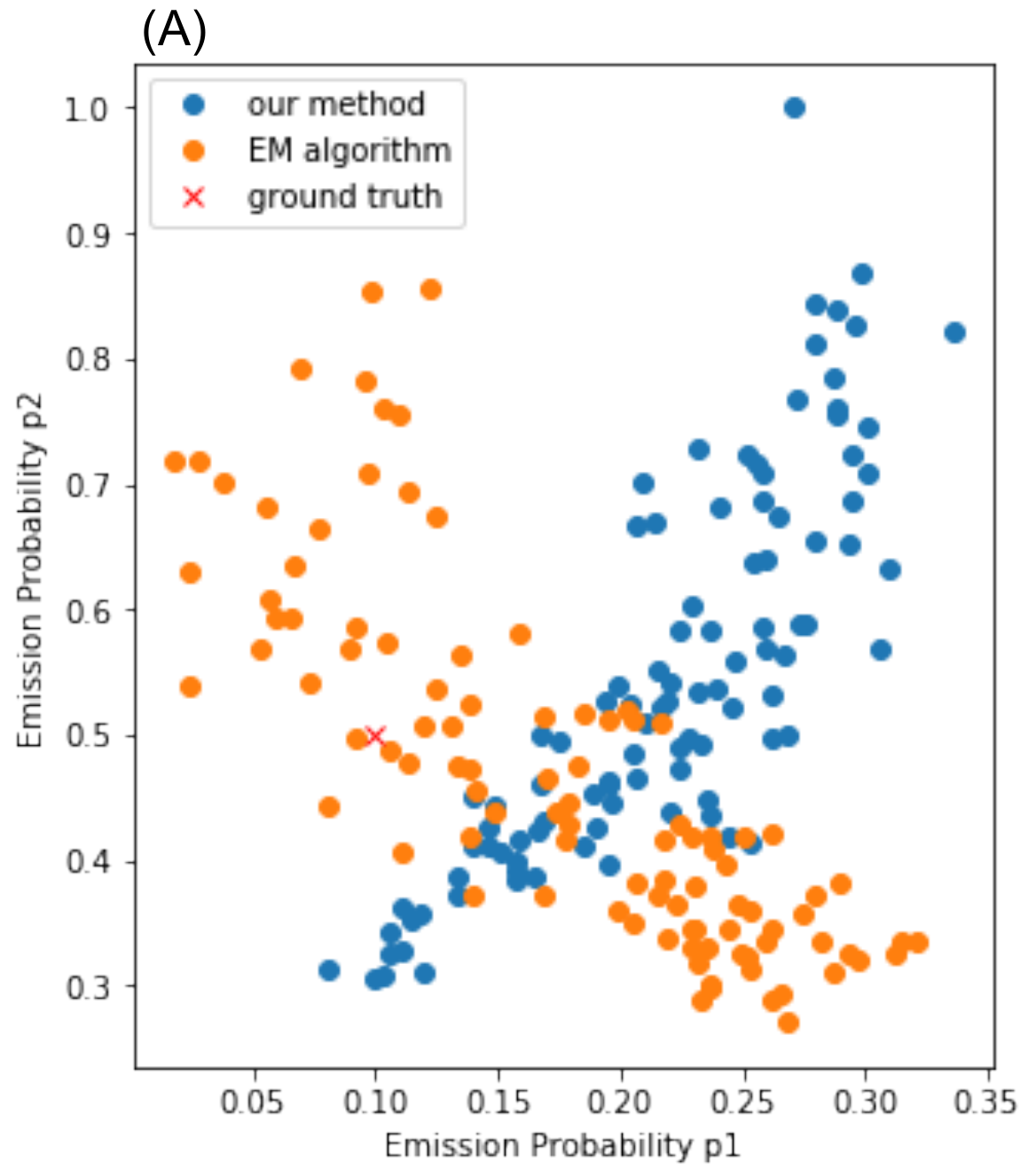}
    \end{subfigure}
    \quad 
    \begin{subfigure}[b]{0.45\textwidth}
        \includegraphics[width=\textwidth]{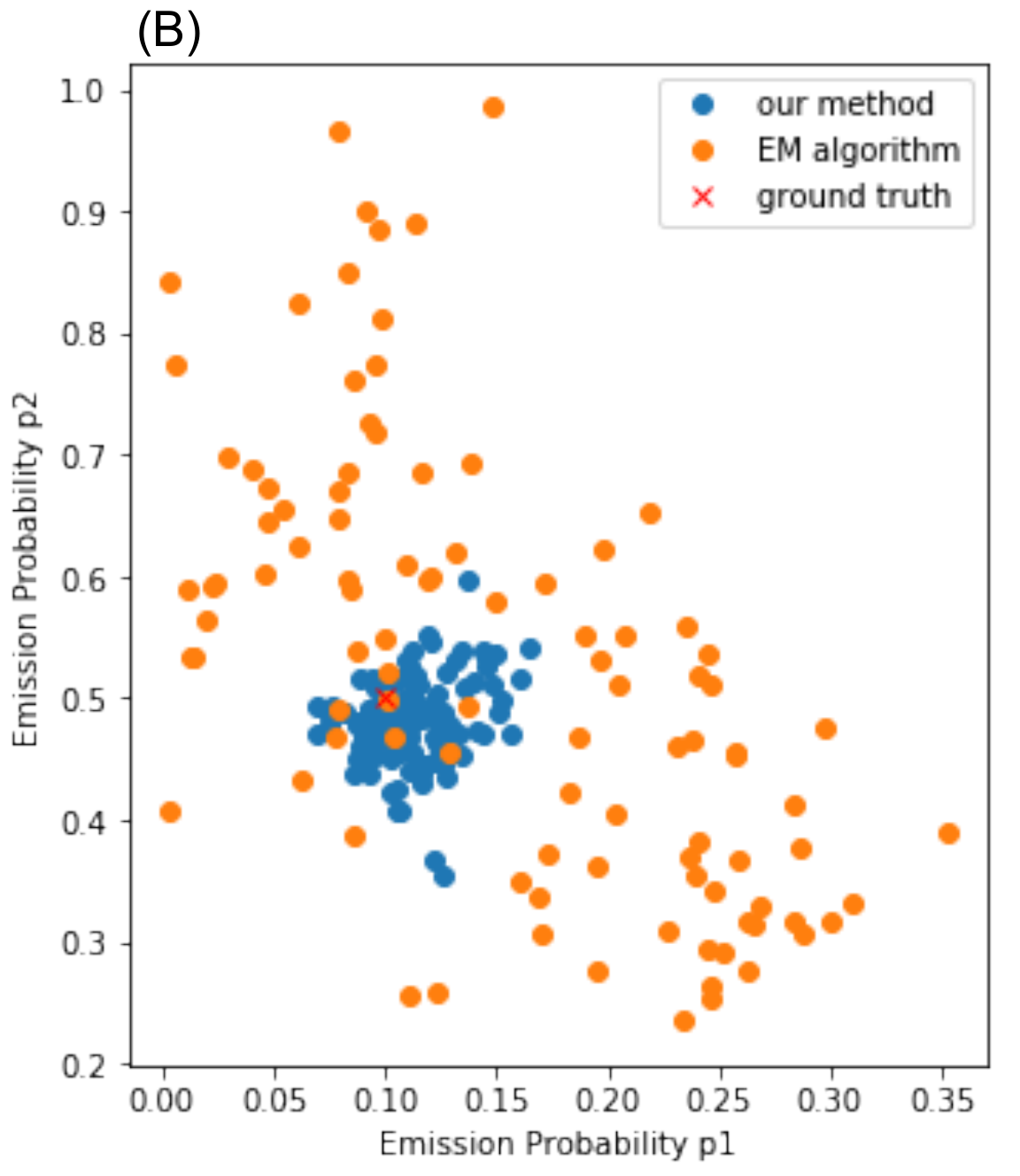}
    \end{subfigure}
\caption{Data is generated under HMM settings with $p_1=0.1$, $p_2=0.5$, and different $p_{12}$. The simulation is repeated for 100 times and each dot is an estimation for $p_1$ and $p_2$ in one realization. (A) $p_{12}=0.1$; mean (std) Euclidean distance is 0.182 (0.089) for our method and 0.163 (0.080) for EM. (B) $p_{12}=0.01$; mean (std) Euclidean distance is 0.041 (0.024) for our method and 0.184 (0.095) for EM.}
\label{fig.emission}
\end{figure}

Apart from decoding, the estimation accuracy of emission probability is also compared for the proposed method and the Baum–Welch's. Following the same simulation above, results of two-dimensional estimation for $p_1$ and $p_2$ are shown in Figure\ref{fig.emission}. 

As $p_{12}$ decreases, the estimated points of the proposed method are closely around the true parameters. Instead, the estimations from the Baum–Welch's are far apart from the truth with choice of different $p_{12}$. Indeed, When $p_{12}=0.01$, the average 2-dim Euclidean distance from the estimations to the true parameter is 0.041 for the proposed method, which is much lower than 0.184 for the Baum–Welch's. 

As a summary in this section, the proposed method has a good performance in both decoding and probability estimation. Though an assumption of independent observations is advocated, the proposed method is robust when a weak serial dependence is present and competitive to HMM even under Markovian conditions.


\section{Encoding-and-Decoding Procedure}

\subsection{The Method}

So far, the choice of threshold in defining an event or large returns plays an importance role. A natural question to ask is how stable is the estimation result with respect to a threshold. For example, an observed return is marked if it is below a threshold $\pi$, the intensity parameter of geometric distribution becomes $p_s(\pi)=F_s(\pi)$ given a hidden state $s$. By assuming the continuity of the underlying distribution $F_s$, $p_s$ is also continuous with $\pi$. Thus, the emission probability under a hidden state would not fluctuate much if $\pi$ varies slightly. Indeed, our experiment shows that the estimated emission probability is not sensitive to $\pi$. To make the notation consistent, we will use $p^{\pi}(t)$ or $F^{\pi}(t)$ if both $t$ and $\pi$ are present.

The idea of dealing with a stochastic process of continuous observations is described as follows. In the encoding phase, we iteratively switch the excursion threshold and discretize the time series into a 0-1 process with each threshold applied. After that, we implement the searching algorithm to decode each process separately. As a consequent result, a vector of estimated emission probability $\hat{p}^{\pi}(t)$ is obtained at time $t$ with different choices of $\pi$. It actually provides an estimation to the Cumulative Distribution Function(CDF) at time $t$ by $\hat{F}^{\pi}(t)=\hat{p}^{\pi}(t)$ where $\hat{F}^{\pi}(t)$ is a function of $\pi$ at a fixed $t$. Take the simulated t-distributed data in Appendix\ref{appendix:simulation} as an example, Figure\ref{fig.eCDFs} shows a series of CDF with a change point embedded in the middle though it is hard to detect by eyes. Lastly, all the decoding information is aggregated in an emission vector $\vec{p}(t)$, and the volatility dynamic is discovered via cluster analysis. Suppose that $\{C^{\pi}(t)\}_t$ is a 0-1 coding sequence obtained by applying a threshold $\pi$ upon the returns, and $\Pi$ is a pre-determined threshold set, for example, $\Pi$ can be a series of quantiles of the marginal distribution $\Pi=\{0.9\,\textit{quantile}, 0.8\,\textit{quantile},..., 0.1\,\textit{quantile}\}$. The encoding-and-decoding algorithm is described in Alg.2.

\noindent\rule{12.5cm}{0.8pt}\\
\textbf{Alg.2} Encoding-and-Decoding\\
\rule{12.5cm}{0.4pt}\\
1. For loop: cycle threshold $\pi$ through $\Pi=\{\pi_1,\pi_2,...,\pi_V\}$:

\quad Define events and code the whole process as a 0-1 digital string $\{C^{\pi}(t)\}_{t=1}^{n}$,
\[C^{\pi}(t) = \begin{cases} 1 & \quad log\frac{X(t)}{X(t-1)} \le \pi \enspace \textit{if} \enspace \pi<0 \enspace \\
1 & \quad log\frac{X(t)}{X(t-1)} \ge \pi \enspace \textit{if} \enspace \pi>0 \\
0 & \quad \textit{otherwise} \end{cases}\]

\quad Repeat step 3 \& 4 in Alg.1 and estimate the probability $\hat{p}^{\pi}(t)$ by (\ref{eq.p}).\\
\hspace{0.5em} End For\\
2. Stack the estimated emission probability at $t$ in a vector $\vec{p}(t)$,
\[
\vec{p}(t):=(\hat{p}^{\pi_1}(t),\hat{p}^{\pi_2}(t),..., \hat{p}^{\pi_V}(t))
\]
3. Merge time points with comparable $\vec{p}(t)$ together via clustering analysis\\
\rule{12.5cm}{0.8pt}

\begin{figure}[h]
\centering
\includegraphics[width=4.1in]{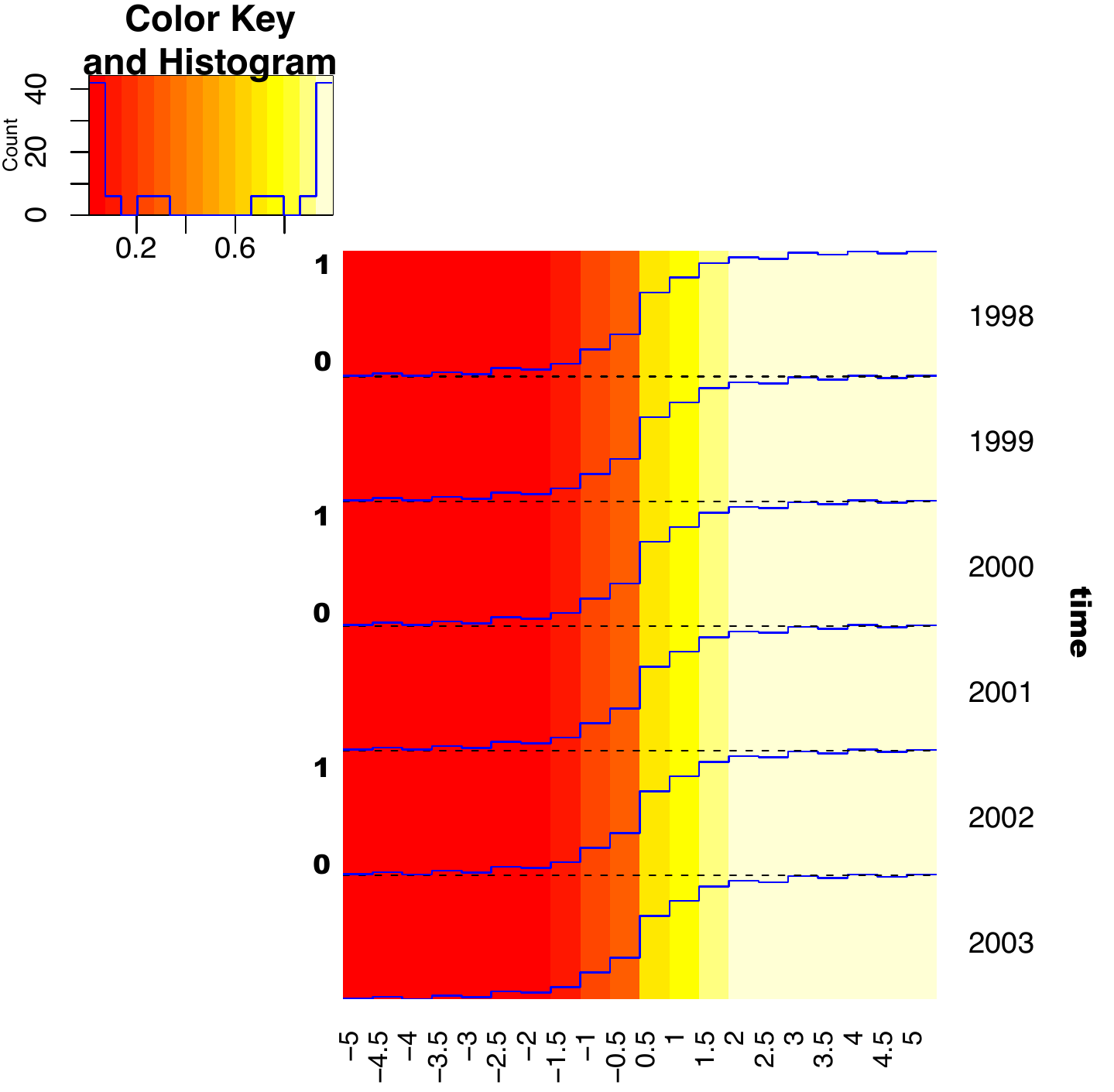}
\caption{Simulated data from 1998 to 2003. Data from 1998 to 2000 follows student-t distribution with degree of freedom 2; data from 2001 to 2003 follows student-t distribution with degree of freedom 5.}
\label{fig.eCDFs}
\end{figure}

It is surprising that how the CDF is estimated based on the only observation at each time stamp. Indeed, the $\hat{F}^{\pi}$ does not consistently converge to the true CDF as the number of thresholds $\pi$ increases due to the limitation of data information. The estimated CDF here depends on the decoding result. Specifically, if there exists a threshold $\pi$ by which the underlying distribution can be separated well, then the decoding achieves a good result to reflect the distributional changes. On the other hand, if the $\pi$ is set not appropriate, for example, the emission probability of the underlying distribution at state $s$ and $s^{'}$ are very close to each other, say $\hat{F}_s^{\pi} \cong \hat{F}_{s^{'}}^{\pi}$, then the decoding algorithm fails to separate the two states with such a threshold applied. There is an ongoing discussion about how to choose a good threshold to discretize a sequence of continuous observations. A heuristic idea is to tune the optional value of $\pi$ such that the estimated probabilities under different states are far apart from each other. For example, consider a max-min estimator of $\pi$,
\begin{equation}
\hat{\pi}=\argmax_{\pi} \min_{s,s^{'}} |\hat{p}_s(\pi)-\hat{p}_{s^{'}}(\pi)|
\end{equation}

It is remarked that the proposed procedure avoids the issue of tuning parameters by imposing a series of thresholds and aggregating all the decoding results together. The information of distributional changes is reserved into the emission vector $\vec{p}(t)$. On the other hand, an irrelevant result with an unreasonable threshold applied would not change the aggregation result much. For example, if $\hat{F}_s^{\pi} \cong \hat{F}_{s^{'}}^{\pi}$, then there is no distributional changes detected in the process, so $\hat{p}^{\pi}(t)$ is a constant for any $t$. In summary, the algorithm is implemented by shifting $\pi$ value from high to low to obtain a sequence of estimated CDFs, although not all the $\pi$'s are meaningful in the decoding. By further combining the estimated emission in a vector, the aggregation sheds a light on differentiating underlying distributions.

\subsection{Clustering Analysis}

\begin{figure}[h]
\centering
\begin{tabular}{@{}c@{}}
    \includegraphics[width=4in]{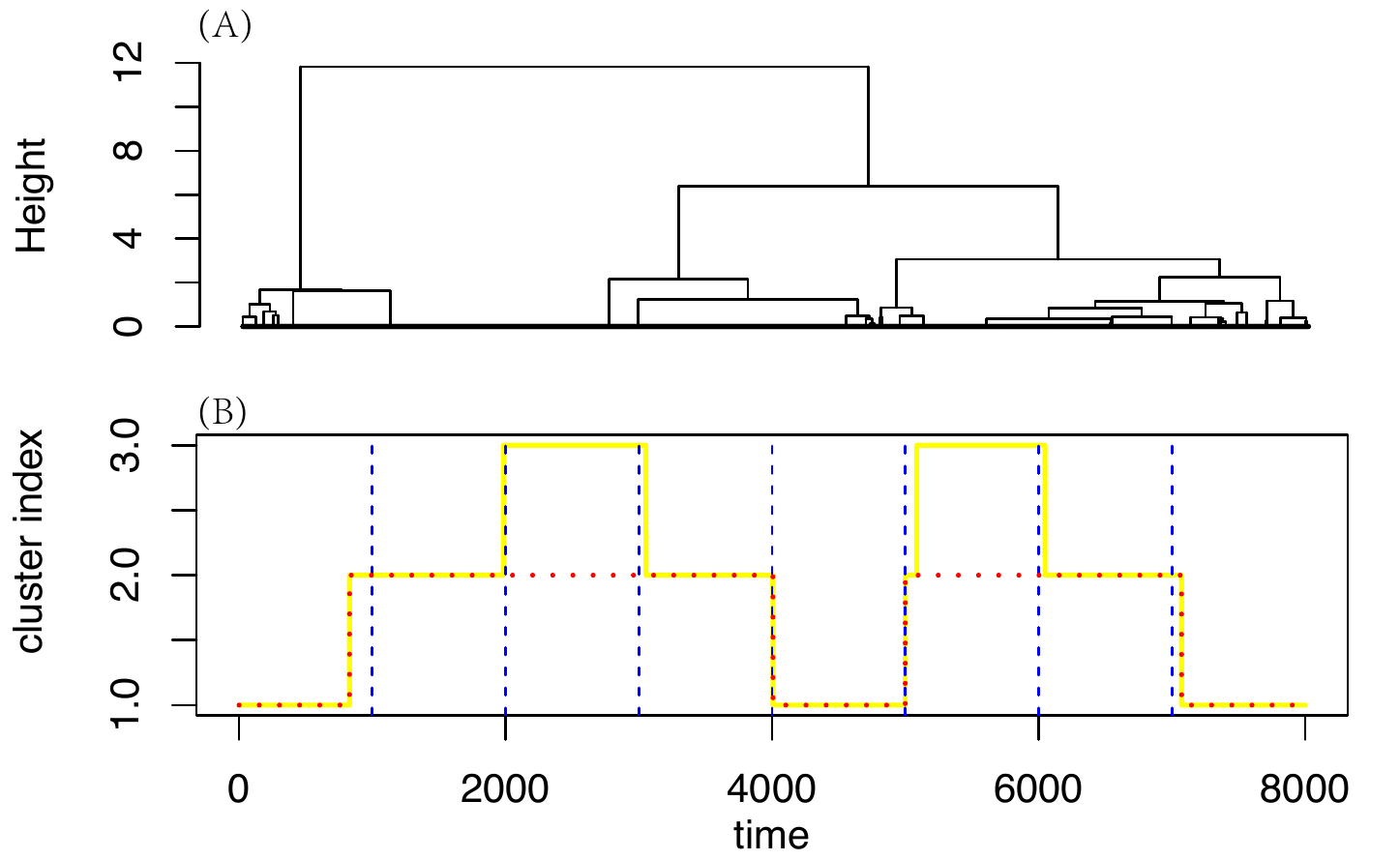}
  \end{tabular}
  \vspace{\floatsep}
  \begin{tabular}{@{}c@{}}
    \includegraphics[width=4in]{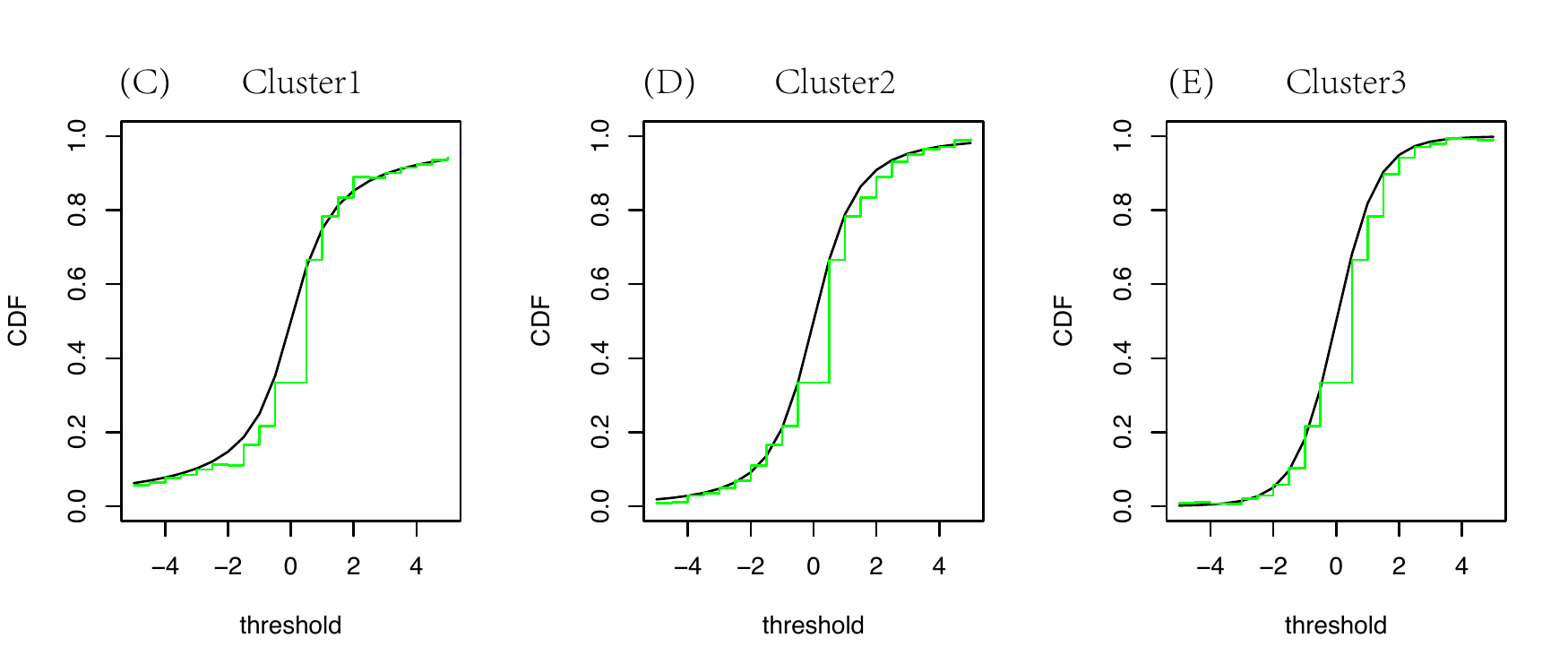}
\end{tabular}
\caption{3-states decoding results for simulated data in Appendix\ref{appendix:simulation}. (A) Hierarchical Clustering Tree; (B) cluster index switching over time; (C),(D),(E): estimated CDF versus true CDF, in cluster 1,2,3, respectively.}
\label{fig.simulation_3statesclustering}
\end{figure}

The next question is how to extract the information in $\{\vec{p}(t)\}_{t=1}^n$, and how many underlying distributions are required to describe the patterns of distribution switching. It actually raises a question for all the regime-switching models to determine the number of underlying states. Generally, the more states are taken into consideration, the less tractable a model becomes. For example, a 2-state lognormal Markov Model contains 6 parameters, while a 3-state model increases the number to 10. The number of states is usually decided subjectively or tuned with an extra criterion. Given the estimated probability vector $\vec{p}(t)$ in Alg.2, the problem above can be resolved by clustering similar time points together such that the CDF curves within each cluster are in a comparable shape. It is proved by \cite{Wang} that the clustering index obtained via hierarchical clustering with `Ward' linkage maintains the asymptotic property in change-point analysis. Furthermore, the number of underlying states can be determined by searching through the number of clusters embedded in the emission vectors $\vec{p}(t)$. Here, hierarchical clustering with `Ward' linkage is implemented to cluster similar time points shown in Figure\ref{fig.simulation_3statesclustering}(A). One can visualize the dendrogram to decide the number of clusters, or employ extra criteria like Silhouette to measure the quality of clustering.

Numerical data simulated with student-t distributions is available in Appendix\ref{appendix:simulation}. The dendrogram shows that 2 or 3 clusters may be embedded inside the observations. If we cut the dendrogram into 3 clusters, the trajectory of cluster indices can almost perfectly represent the alternating hidden states, see Figure\ref{fig.simulation_3statesclustering}(B). If 2 is taken rather than 3, the result makes sense as well since cluster2 and cluster3 are combined together as a contradiction to the high-intensity cluster or cluster1. By calculating the average emission curve in each cluster, one can compare the estimated probability function with the theoretical distributions. It shows in Figure\ref{fig.simulation_3statesclustering} (C)-(E) that the emission curve is a goodness-of-fit in each cluster.

It is demonstrated that the decoding result is robust to the number of hidden states that is supposed in each 0-1 process. Without the prior knowledge of 3-states embedding, if we implement a 2-states or 4-states decoding schedule, the clustering trajectory can still describe the distributional changes well, see Figure\ref{fig.simulation_2statesclustering} and Figure\ref{fig.simulation_4statesclustering} in Appendix\ref{appendix:graph}.

\subsection{Simulation: Continuous Observations}

In this section, we present simulation results for the encoding-and-decoding approach under continuous Hidden Markov Model (HMM) settings. Decoding errors are compared with well-known parametric approaches. All the computation of parametric approaches is completed using the Python package \textit{hmmlearn}.

In the first setting, data is generated by Gaussian HMM with 2 hidden states $S_1$ and $S_2$. We set both conditional means as $\mu_1=\mu_2=0$ and conditional variance as $\sigma^2_1$ and $\sigma^2_2$. Gaussian HMM with 2 hidden states is implemented to calibrate model parameters and decode hidden states. In the second setting, data is generated under Gaussian Mixture Model HMM (GMMHMM) with 2 hidden states and 2 Gaussian components, $w_a \mathcal{N}(0,\sigma^2_{i,a}) + w_b \mathcal{N}(0,\sigma^2_{i,b})$, in which $w_a$ is the weight of component `a' and $\sigma^2_{i,a}$ is the variance of component `a' under state $S_i$, for $i=1,2$. GMMHMM with 2 hidden states and 2 Gaussian components is implemented for comparison. 

The mean of decoding error rates under both settings is reported in Table\ref{tab.GHMM} and Table\ref{tab.GMMHMM}, respectively. With a fixed sample size $n=1000$, all the decoding result gets better as the transition probability $p_{12}$ decreases. Especially, as the serial dependence is low enough or $p_{12}$ is less than 0.01, the proposed method is far better than the parametric models from which the data is generated. The small sample size may explain the failure of parametric models. It shows that the proposed non-parametric method is more stable and reliable than the parametric when data resource is limited in a real application.

\begin{table}[h] 
\centering
\caption{Gaussian HMM setting: average decoding error rates; better error rate marked in bold}
\scalebox{0.85}{%
\begin{tabular}{l|cc|cc|cc|}
\cline{2-7}
                               & \multicolumn{2}{c|}{$\sigma^2_1$=0.4,   $\sigma^2_2$=1} & \multicolumn{2}{c|}{$\sigma^2_1$=1,   $\sigma^2_2$=2} & \multicolumn{2}{c|}{$\sigma^2_1$=1,   $\sigma^2_2$=3} \\ \hline
\multicolumn{1}{|l|}{$p_{12}$} & GaussianHMM                & Our Method                 & GaussianHMM             & Our Method                  & GaussianHMM               & Our Method                \\ \hline
\multicolumn{1}{|l|}{0.1}      & \textbf{0.4067}            & 0.4576                     & 0.4592                  & \textbf{0.4589}             & \textbf{0.3220}           & 0.4616                    \\
\multicolumn{1}{|l|}{0.05}     & \textbf{0.3631}            & 0.4321                     & 0.4586                  & \textbf{0.4462}             & \textbf{0.2623}           & 0.4272                    \\
\multicolumn{1}{|l|}{0.01}     & 0.3558                     & \textbf{0.2771}            & 0.4314                  & \textbf{0.3330}             & \textbf{0.1755}           & 0.2604                    \\
\multicolumn{1}{|l|}{0.005}    & 0.3447                     & \textbf{0.2160}            & 0.4298                  & \textbf{0.2488}             & 0.1842                    & \textbf{0.1609}           \\ \hline
\end{tabular}
}
\label{tab.GHMM}
\end{table}

\begin{table}[h]
\centering
\caption{GMMHMM setting: average decoding error rates; better error rate marked in bold}
\scalebox{0.75}{%
\begin{tabular}{l|cc|cc|cc|cc|}
\cline{2-9}
\multicolumn{1}{c|}{\multirow{2}{*}{}} & \multicolumn{4}{c|}{$\sigma^2_{1,a}$=0.1 ,\quad $\sigma^2_{1,b}$=0.5 ,\quad $\sigma^2_{2,a}$=1 ,\quad $\sigma^2_{2,b}$=1.5}                            & \multicolumn{4}{c|}{$\sigma^2_{1,a}$=0.1 ,\quad $\sigma^2_{1,b}$=0.8 ,\quad $\sigma^2_{2,a}$=0.5 ,\quad $\sigma^2_{2,b}$=1.5}                           \\ \cline{2-9} 
\multicolumn{1}{c|}{}                  & \multicolumn{2}{c|}{$w_a$=$w_b$=0.5} & \multicolumn{2}{c|}{$w_a$=0.3 , $w_b$=0.7} & \multicolumn{2}{c|}{$w_a$=$w_b$=0.5} & \multicolumn{2}{c|}{$w_a$=0.3 , $w_b$=0.7} \\ \hline
\multicolumn{1}{|l|}{$p_{12}$}              & GMMHMM            & Our Method       & GMMHMM       & Our Method           & GMMHMM       & Our Method            & GMMHMM       & Our Method            \\ \hline
\multicolumn{1}{|l|}{0.1}              & \textbf{0.3943}   & 0.4533           & 0.4326       & 0.4622               & 0.4370       & 0.4573                & 0.4661       & \textbf{0.4596}       \\
\multicolumn{1}{|l|}{0.05}             & \textbf{0.3661}   & 0.4137           & 0.4135       & 0.4150               & 0.4368       & \textbf{0.4259}       & 0.4605       & \textbf{0.4401}       \\
\multicolumn{1}{|l|}{0.01}             & 0.3456            & \textbf{0.2342}  & 0.4072       & \textbf{0.2287}      & 0.4290       & \textbf{0.3390}       & 0.4565       & \textbf{0.3663}       \\
\multicolumn{1}{|l|}{0.005}            & 0.3292            & \textbf{0.1431}  & 0.3810       & \textbf{0.1515}      & 0.4236       & \textbf{0.2558}       & 0.4550       & \textbf{0.2981}       \\ \hline
\end{tabular}
}
\label{tab.GMMHMM}
\end{table}

\section{Real Data Experiments}

\subsection{Forecasting}

Forecasting stock prices or returns has become one of the basic topics in financial markets. As one of the forecasting models, continuous Hidden Markov Model has been widely implemented due to its strong statistical foundation and tractability. Based on the work of \cite{Hassan, Nguyen}, the 1-step-ahead forecasting can be done by looking for a ``similar'' historical data set that is a close match to the current values. In this section, we implement our encoding-and-decoding approach under the forecasting framework of HMM to predict stock returns and compare the results with that of Gaussian HMM.

We first introduce the stock forecasting process using HMM by \cite{Nguyen}. Given observations $\{Y(t)\}_{t=1}^T$, suppose that the goal is to predict stock price at time $T+1$. In the first step, a fixed training window of length $D$ is chosen and then training data set $Y_{tr}=\{Y(t)\}_{t=T-D+1}^T$ is used to calibrate HMM's parameters denoted by $\lambda$ and calculate observation probability $P(Y_{tr}|\lambda)$ given $\lambda$ . In the second step, we move the training window backward by one stamp to obtain $Y_{-1}=\{Y(t)\}_{t=T-D}^{T-1}$, and keep moving backward stamp by stamp until all the historical data is covered, so $Y_{-k}=\{Y(t)\}_{t=T-D+1-k}^{T-k}$, for $k=1,...,T-D$. In the last step, we find a data set $Y_{-k^*}$ which is the best match to the training data, i.e. $P(Y_{-k^*}|\lambda) \cong P(Y_{tr}|\lambda)$. Thus, $\hat{Y}(T+1)$ is predicted by
\begin{equation}
Y(T)+ (Y(T-k^*+1) - Y(T-k^*)) \times sign(P(Y_{tr}|\lambda)-P(Y_{-k^*}|\lambda))
\end{equation}
Similarly, to predict the stock price at time $T+2$, the training window moves one stamp forward, so the training data is updated by $Y_{tr}=\{Y(t)\}_{t=T-D+2}^{T+1}$.

\begin{figure}[h]
\centering
\begin{tabular}{@{}c@{}}
    \includegraphics[width=4.5in]{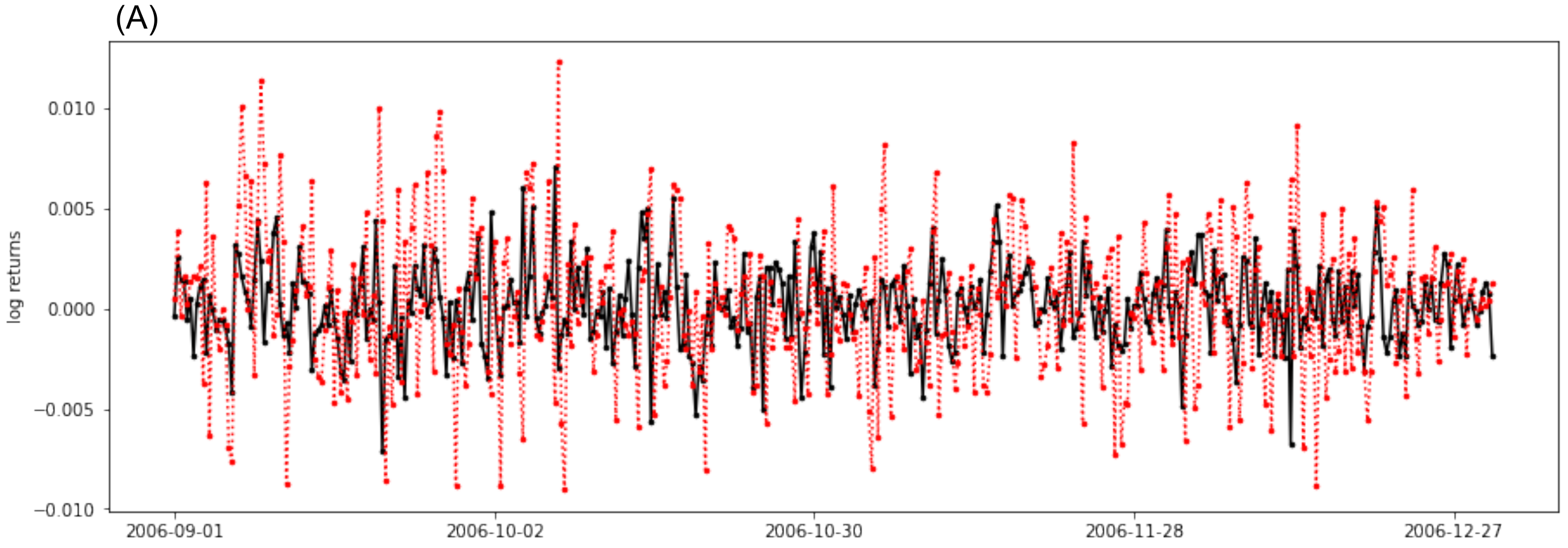}
  \end{tabular}
  \vspace{\floatsep}
  \begin{tabular}{@{}c@{}}
    \includegraphics[width=4.5in]{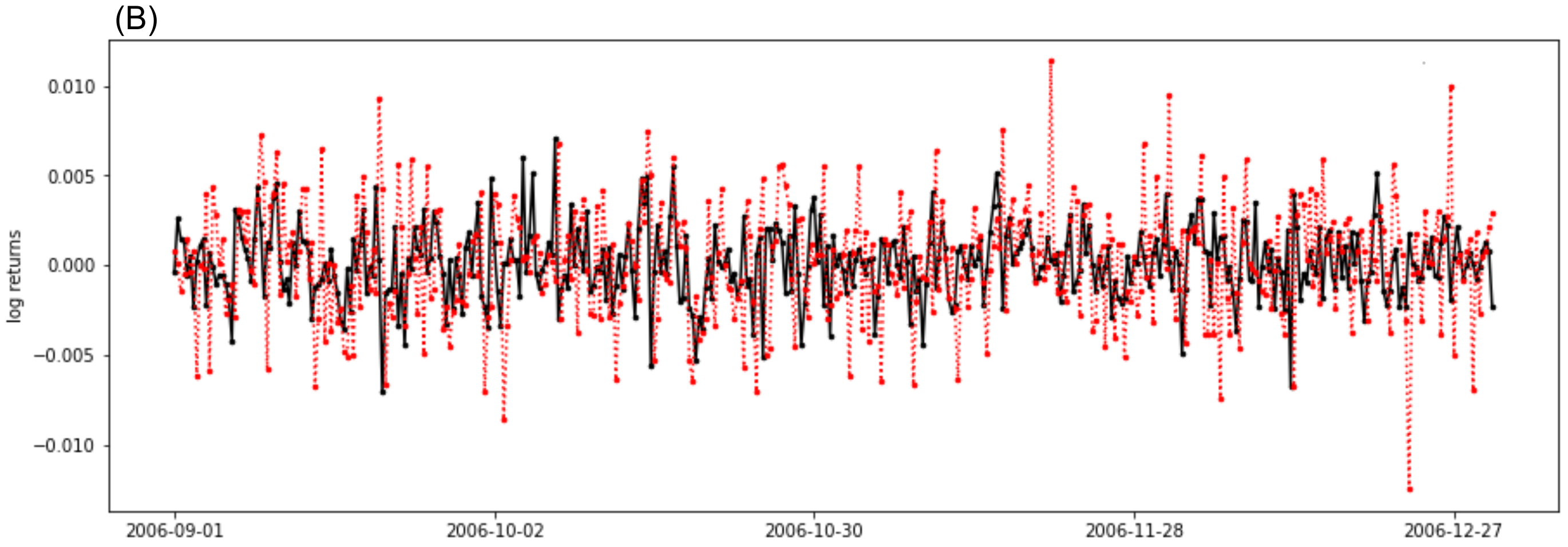}
\end{tabular}
\caption{Time plot of IBM hourly returns from from September to December, 2006. Solid line is real return value and dashed line 1-step-ahead forecasts via (A) Gaussian Hidden Markov Model (B) the proposed method. The $D$ value is set at 200.}
\label{fig.forecasting}
\end{figure}

In the implementation of Gaussian HMM, model parameters are summarized by $\lambda:=\{\pi,A,\mu|s,\sigma|s\}$ where $\mu|s$ and $\sigma|s$ are mean and standard deviation of Gaussian distribution given hidden state $s$, respectively. Once $\lambda$ is estimated in the training set, the probability of observations can be calculated by the forward-backward algorithm. However, the Gaussian assumption is usually violated in describing the heavy-tail returns, which makes the forecasting model implausible. While, the proposed encoding-and-decoding approach is able to estimate the probability of observations without any supposed distribution family. By applying a series of quantile threshold $\pi=\{\pi_1,\pi_2,...,\pi_V\}$ in encoding, one can figure out the bins in which $Y(t)$ is located, so the probability of observing $Y(t)$ can be calculated based on the estimated emission $\hat{F}^{\pi}$ by
\begin{equation}
\hat{P}(Y(t)|S)=\sum_{i=1}^{V+1}(\hat{F}^{\pi_i}-\hat{F}^{\pi_{i-1}})\mathbf{1}_{\{\pi_i< Y(t)\le \pi_{i-1}\}}
\end{equation}
where $\hat{F}^{\pi_0}$ is denoted as 0 and $\hat{F}^{\pi_{V+1}}$ as 1. The probability of a set of observations is finally calculated due to the independence assumption, so
\begin{equation}
\hat{P}(Y_{-k}|S)=\prod_{t=T-D+1-k}^{T-k} \hat{P}(Y(t)|S)
\end{equation}

In the real data experiment, hourly returns from January to August 2006 is trained to predict its next-hour-ahead returns from September to December 2006. Gaussian HMM with 4 hidden states is implemented for comparison. Since the choice of window length $D$ is not well defined, two values are tried by $D=100,200$. To measure the model performance, three error estimators are reported for four technology stocks in Table\ref{tab.forecasting}- root mean squared error(RMSE), mean absolute error(MAE), and mean absolute percentage error(MAPE). Figure\ref{fig.forecasting} visualizes the forecasting result for stock of IBM. In the forecasting task, the proposed method slightly outperforms the classic Gaussian HMM in most cases except `INTC' at $D=100$. It can be explained according to our previous experiments that the proposed method takes advantage of decoding hidden states and estimating emission probability compared with HMM. The violation of the Gaussian assumption in HMM may restrict its effectiveness in the real application.

\begin{table}[h]
\centering
\caption{1-step-ahead prediction error for four different stocks- IBM Common Stock(IBM), Intel Corporation(INTC), NVIDIA Corporation(NVDA), and Broadcom Inc(BRCM); better result marked in bold}
\scalebox{0.85}{%
\begin{tabular}{|c|c|cc|cc|}
\hline
                        & {}                            & \multicolumn{2}{c|}{D=100}        & \multicolumn{2}{c|}{D=200}       \\ \cline{3-6} 
\multirow{-2}{*}{Stock} & \multirow{-2}{*}{{Criterion}} & GaussianHMM     & Our Method      & GaussianHMM    & Our Method      \\ \hline
                        & RMSE                                               & 0.0041          & \textbf{0.0040} & 0.0041         & \textbf{0.0038} \\
                        & MAE                                                & 0.0032          & \textbf{0.0031} & 0.0032         & \textbf{0.0029} \\
\multirow{-3}{*}{IBM}   & MAPE                                               & \textbf{6.487}  & 6.986           & \textbf{6.366} & 6.502           \\ \hline
                        & RMSE                                               & \textbf{0.0057} & 0.0062          & 0.0067         & \textbf{0.0058} \\
                        & MAE                                                & \textbf{0.0044} & 0.0046          & 0.0050         & \textbf{0.0045} \\
\multirow{-3}{*}{INTC}  & MAPE                                               & \textbf{5.046}  & 6.656           & 7.402          & \textbf{6.728}  \\ \hline
                        & RMSE                                               & 0.0112          & \textbf{0.0111} & 0.0126         & \textbf{0.0114} \\
                        & MAE                                                & 0.0090          & \textbf{0.0088} & 0.0096         & \textbf{0.0089} \\
\multirow{-3}{*}{NVDA}  & MAPE                                               & 10.122          & \textbf{8.035}  & 8.750          & \textbf{8.166}  \\ \hline
                        & RMSE                                               & 0.0117          & \textbf{0.0109} & 0.0118         & \textbf{0.0113} \\
                        & MAE                                                & 0.0091          & \textbf{0.0086} & 0.0088         & \textbf{0.0086} \\
\multirow{-3}{*}{BRCM}  & MAPE                                               & 13.905          & \textbf{8.246}  & \textbf{9.550} & 9.715           \\ \hline
\end{tabular}
}
\label{tab.forecasting}
\end{table}

\subsection{Volatility Dynamics in High-frequent Data}

In this section, the most high-frequent tick-by-tick data is analyzed for S$\&$P500. The stock returns are calculated in a market time scale which is measured by transaction rather than the real-time clock. The analysis was firstly suggested by \cite{14}, and then worked thoroughly by \cite{15}. A well-known example is a random-walk model suggesting that the variance of returns depends on the number of transactions. Following the idea above, we apply the tiniest sampling rate to alleviate the serial dependency. It is reasonable to assume that the stock returns are exchangeable within a certain number of transactions.

\begin{figure}[!h]
\centering
\includegraphics[width=3.4in]{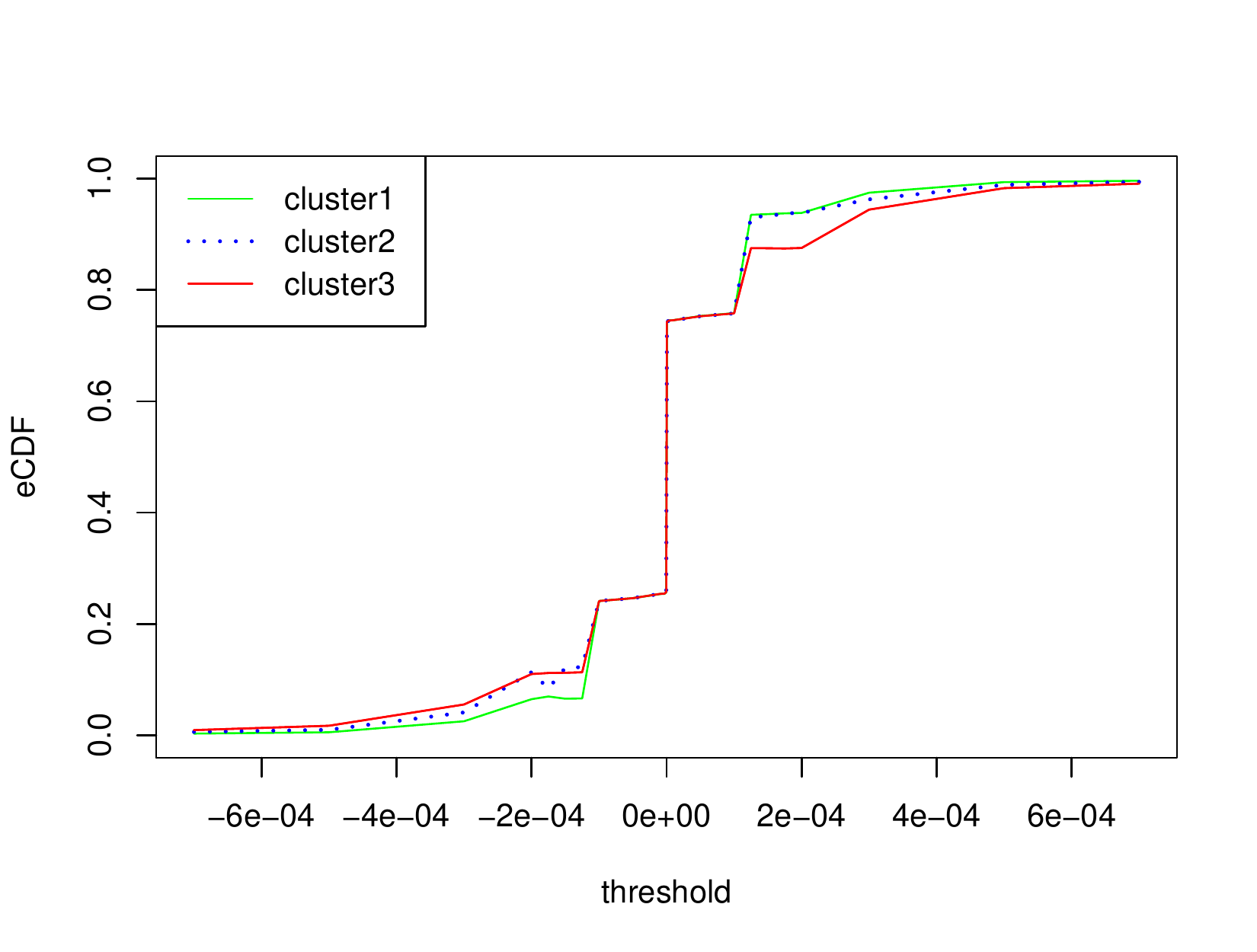}
\caption{The estimated CDFs for the 3 clusters of IBM in January 2006.}
\label{fig.IBM_eCDFs}
\end{figure}

The encoding-and-decoding algorithm is implemented to discover the volatility dynamics for every single stock in S\&P500. Since the result is not sensitive to the number of hidden states, a 2-state decoding procedure is applied and the number of clusters is determined according to the tree height of hierarchical clustering. It turns out that there are 3 potential clusters embedded in the returns of IBM in 2006. The estimated CDF given each cluster is shown in Figure\ref{fig.IBM_eCDFs}. The heavier-tail distribution of cluster3 reflects a high-volatility phase; cluster1 indicates a phase with low volatility. As a phase in the middle, cluster2 shows an asymmetric distribution with a heavy tail on the left but a light tail on the right. Instead, cluster1 and cluster3 look more symmetric on both sides.

\begin{figure}[!h]
\centering
\includegraphics[width=4in]{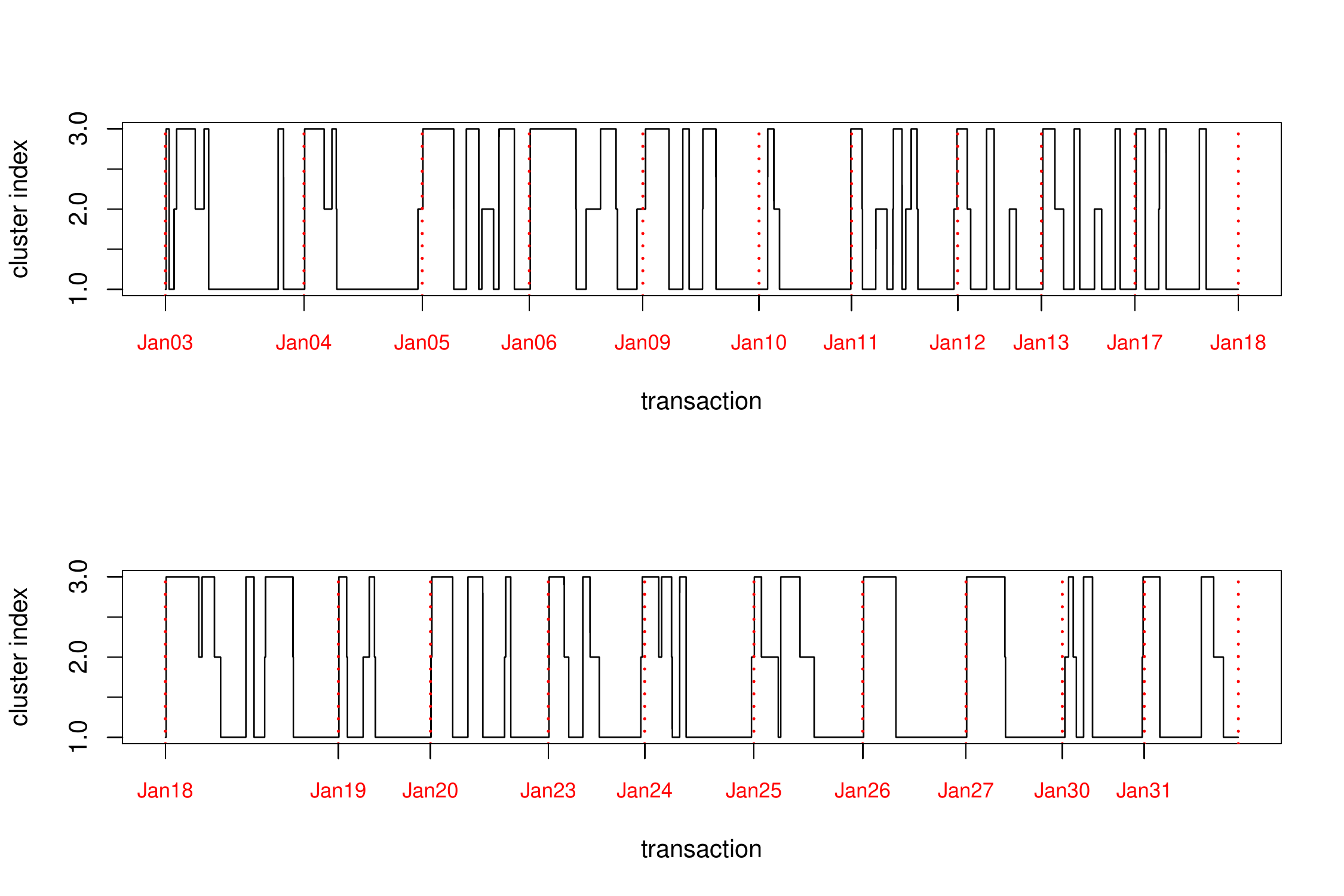}
\caption{Recovered volatility trajectory of IBM in January 2006.}
\label{fig.IBM_parttern}
\end{figure}

The single-stock volatility dynamic is then present by cluster index returned via Alg.2. The dynamic pattern of IBM in January 2006 is shown in Figure\ref{fig.IBM_parttern}. According to the previous notation, cluster1, cluster2, and cluster3 indicates a low-, median-, and high-volatility phase, respectively. Based on the daily segments, it is clear that the unstable high-volatility mostly appears at the beginning of a stock market, and usually shows up twice or three times per day.

\subsection{Information Flows between Stocks}

Beyond detecting the volatility dynamics for a single stock, we further consider the network among all the S$\&$P500 to present how one stock's returns is related to that of another. Such a relationship can be quantified by calculating the cross-correlation between two stock time series \cite{23,24} and correlation matrix was investigated via random matrix theory \cite{25} or clustering analysis \cite{26}. Conditional correlation \cite{27} and partial correlation \cite{28} were studied to provide information about how the relationship of two stocks is eventually influenced given other stocks. However, since the empirical distribution of returns is very different from Gaussian and correlation is only adaptive for a linear relationship, a distribution-free and non-linear measurement is studied to measure the financial connection. Transfer Entropy(TE) \cite{17,18}, as an extension of the concept of Granger causality \cite{16}, was proposed to measure the reduction of Shannon's entropy in forecasting a target variable via the past value of a source variable. Denote the target variable at time $t$ as $Y_t$ and the source variable at $t$ as $X_t$. The Transfer Entropy from $X$ to $Y$ in terms of past $l$ lags is defined by,
\[TE_{X\rightarrow Y}(l)=\sum_{t=l+1}^n P(Y_t,Y_{(t-l):(t-1)},X_{(t-l):(t-1)})log\frac{P(Y_t | Y_{(t-l):(t-1)},X_{(t-l):(t-1)})}{P(Y_t | Y_{(t-l):(t-1)})}\]
It is remarked that the measure is asymmetric, generally, $TE_{X\rightarrow Y} \ne TE_{Y\rightarrow X}$.

However, it is computationally infeasible to calculate the exact TE value due to the difficulty in estimating a conditional distribution or joint distribution especially when $l$ is large. In the application of finance, people commonly cut the observation range into disjoint bins and assign a binning symbol to each data point \cite{30,31,32}. However, a straightforward implementation of binning with equal probability for every symbol will lead to sensible results \cite{29}. To the best of our knowledge, it still lacks in the literature to digitize the stock returns and effectively reveal the dynamic volatility. The simple binning methods such as histogram or clustering fail to catch the excursion of large returns, so only the trend of returns is studied but with dynamic pattern or volatility missing. The encoding-and-decoding procedure remedies the drawbacks of simple binnings by calculating TE through recovered volatility states.
The measure of information flow is improved in measuring the causality of stock volatility rather than the similarity of trend patterns.

\begin{figure}[h]
\centering
\begin{tabular}{@{}c@{}}
    \includegraphics[width=.47\linewidth]{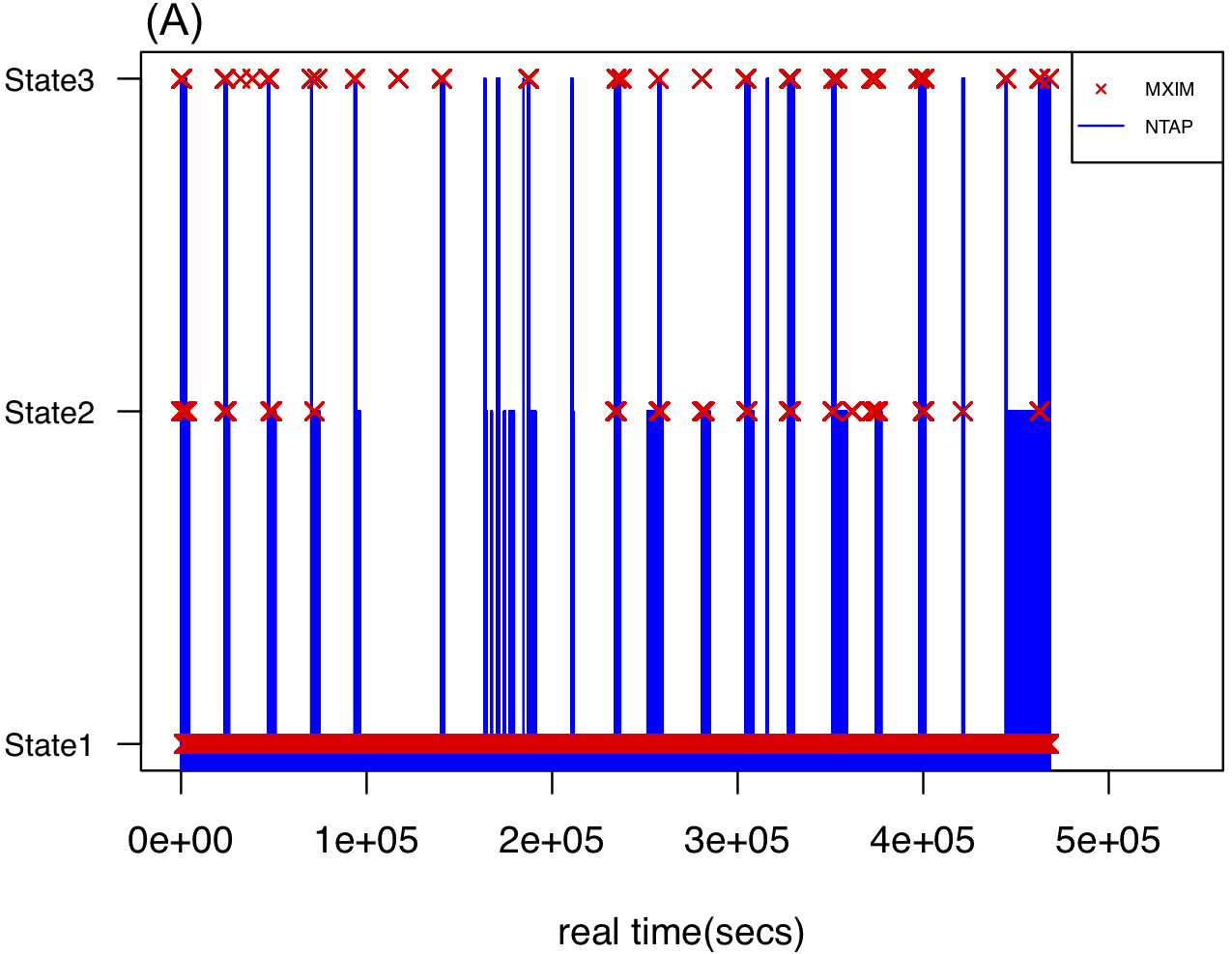}
  \end{tabular}
  \vspace{\floatsep}
  \begin{tabular}{@{}c@{}}
    \includegraphics[width=.47\linewidth]{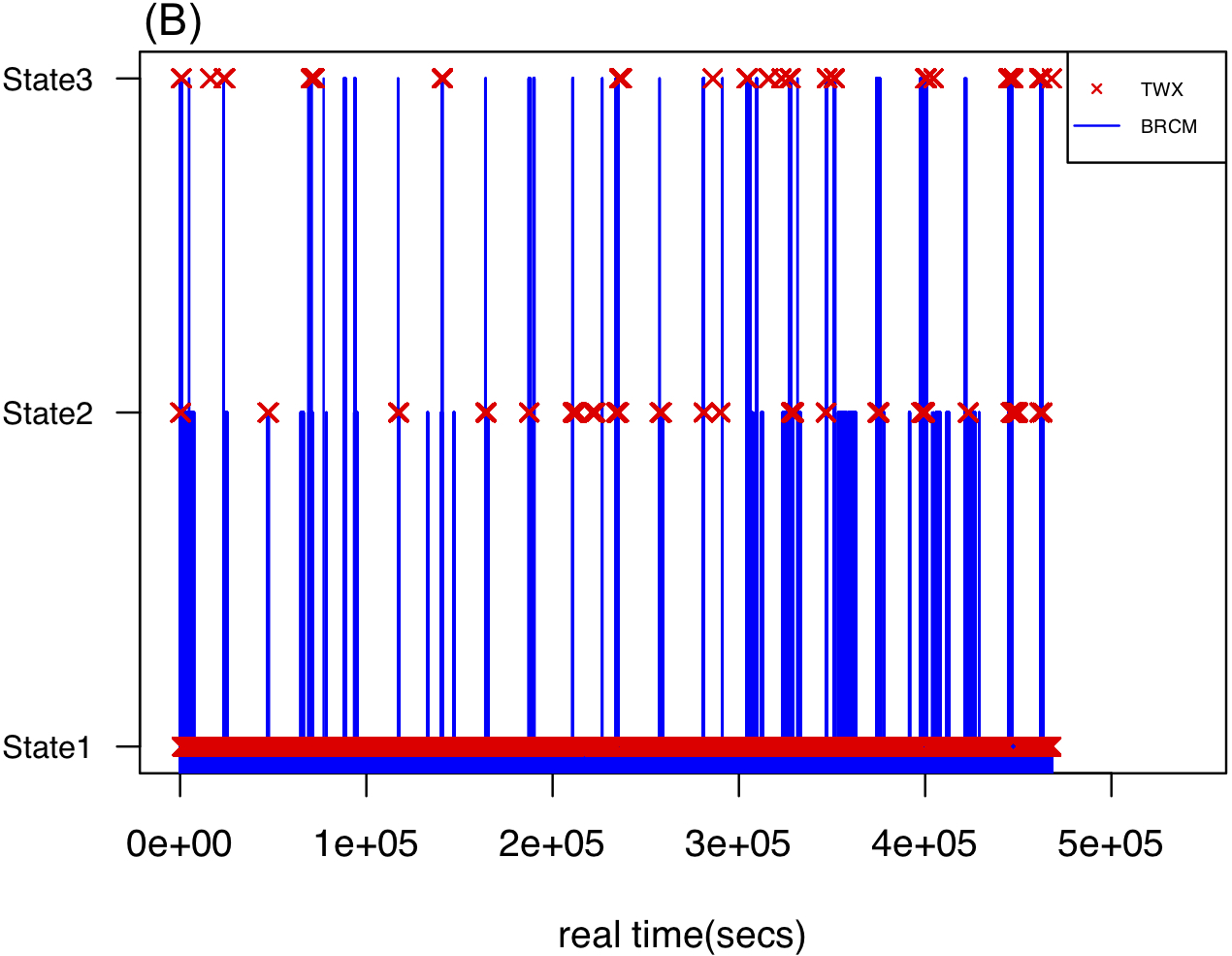}
\end{tabular}
\caption{A pair of volatility trajectories summarized in real time. (A) MXIM(Maxim Integrated Products, Inc.) v.s NTAP(NetApp, Inc.); (B) TWX(Time Warner, Inc.) v.s BRCM(Broadcom, Inc.)}
\label{fig.associated_pair}
\end{figure}

To deal with missing data in high-frequency trading time series, a transformation is proposed to link a pair of stocks into the same time scale, then a pairwise dependency is measured based on Transfer Entropy. The detail of implementation is shown in Appendix\ref{appendix:TE}. A pair of decoded symbolic sequences are shown in Figure\ref{fig.associated_pair}. The higher information flow from $X$ to $Y$, the stronger impact that $X$ promotes volatility of $Y$. In the first example, Figure\ref{fig.associated_pair}(A) shows that MXIM and NTAP share a large intersection in volatility phases. Especially, when MXIM is in volatility, the price of NTAP has a high probability to be in state3. The information flow from MXIM to NTAP is 0.039 and 0.016 in reverse. In the second example, Figure\ref{fig.associated_pair}(B) shows that TWX has a stronger influence on the volatility stages of BRCM. The measure from TWX to BRCM is 0.036 and 0.026 in reverse.


Once the Transfer Entropy is calculated for all pairs of stocks in S$\&$P500, the result is recorded in a $500 \times 500$ asymmetric matrix with the entry value of the $i$-th row and the $j$-column as the information flow from the $i$-th stock to the $j$-th stock. We rearrange the rows and columns such that the sum of rows and the sums of columns are in ascending order, respectively. The reordered TE matrix is shown in Figure\ref{fig.TEmatrix} in Appendix\ref{appendix:graph}. The idea of reordering follows the discussion about the node centrality for directed networks in \cite{31}. Two types of node strength are considered for incoming and outgoing edges. An incoming node strength at node $i$, denoted as $NS^{i}_{in}$, is defined by the sum of the weights of all the incoming edges to $i$,
\begin{equation}
NS^{i}_{in} = \sum_{j=1}^n TE_{j\rightarrow i}
\end{equation}
Similarly, an outgoing node strength, denoted as $NS^{i}_{out}$, is defined by the sum of the weights of all the outgoing edges from $i$,
\begin{equation}
NS^{i}_{out} = \sum_{j=1}^n TE_{i\rightarrow j}
\end{equation}
If a stock has a large value of incoming node strength, it receives more information flow, which means the stock is strongly influenced by others; while, if a stock has a large outgoing node strength, it sends more impacts to other stocks. The top30 stocks with the largest incoming and outgoing node strength values are reported in Table\ref{tab.top30} in Appendix\ref{appendix:graph}. If we take the intersection between the top30 incoming nodes and the top30 outgoing nodes, a group of most central stocks gets returned. The central stocks can be interpreted as intermediate nodes connecting all the other stocks in the S$\&$P500 network. The central stocks include CHK(Chesapeake Energy), VLO(Valero Energy), NTAP(NetApp, Inc.), BRCM(Broadcom, Inc.), and TWX(Time Warner, Inc.), which are all big manufacturers, retailers, suppliers, or media covering the important fields in the United States.

\subsection{S\&P500 Networks}

\begin{figure}[!h]
\centering
\includegraphics[width=3.2in]{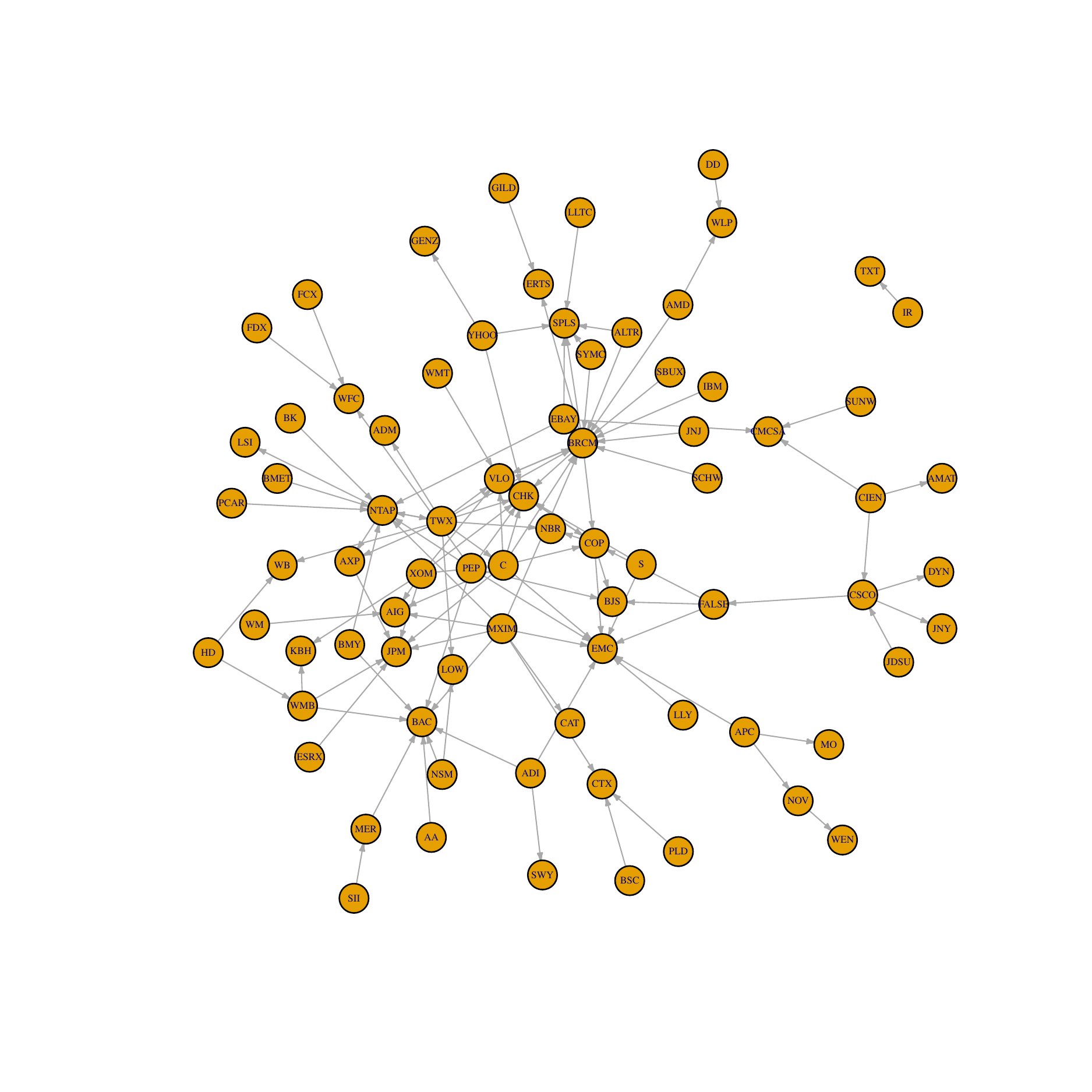}
\caption{A directed network of S$\&$P500: edges with the strongest weights and the conjunct nodes are shown.}
\label{fig.network01}
\end{figure}

In this subsection, we present two different types of networks to illustrate the volatility connection among the S$\&$P500 in 2006. A weighted directed network is established by regarding each stock as a node, information flow from one node to another as an edge, and the Transfer Entropy value as the weight of an edge. Nodes with weak dependency are filtered out, so only the strongest edges and their conjunct nodes are shown in Figure\ref{fig.network01}. Apart from the central stocks such as CHK, VLO, NTAP, and BRCM, the result shows that big investment corporations, such as JPM(JPMorgan), BAC(Bank of America), and C(Citigroup) also heavily depend on other stocks. Instead, TWX(Time Warner, Inc.), MXIM(Maxim Integrated Products Inc.), APC(Apple inc.), EBAY(eBay Inc.), and YHOO(Yahoo! Inc.) has a primary impact on other S$\&$P500.

\begin{figure}[!h]
\centering
\begin{tabular}{@{}c@{}}
    \includegraphics[width=.47\linewidth]{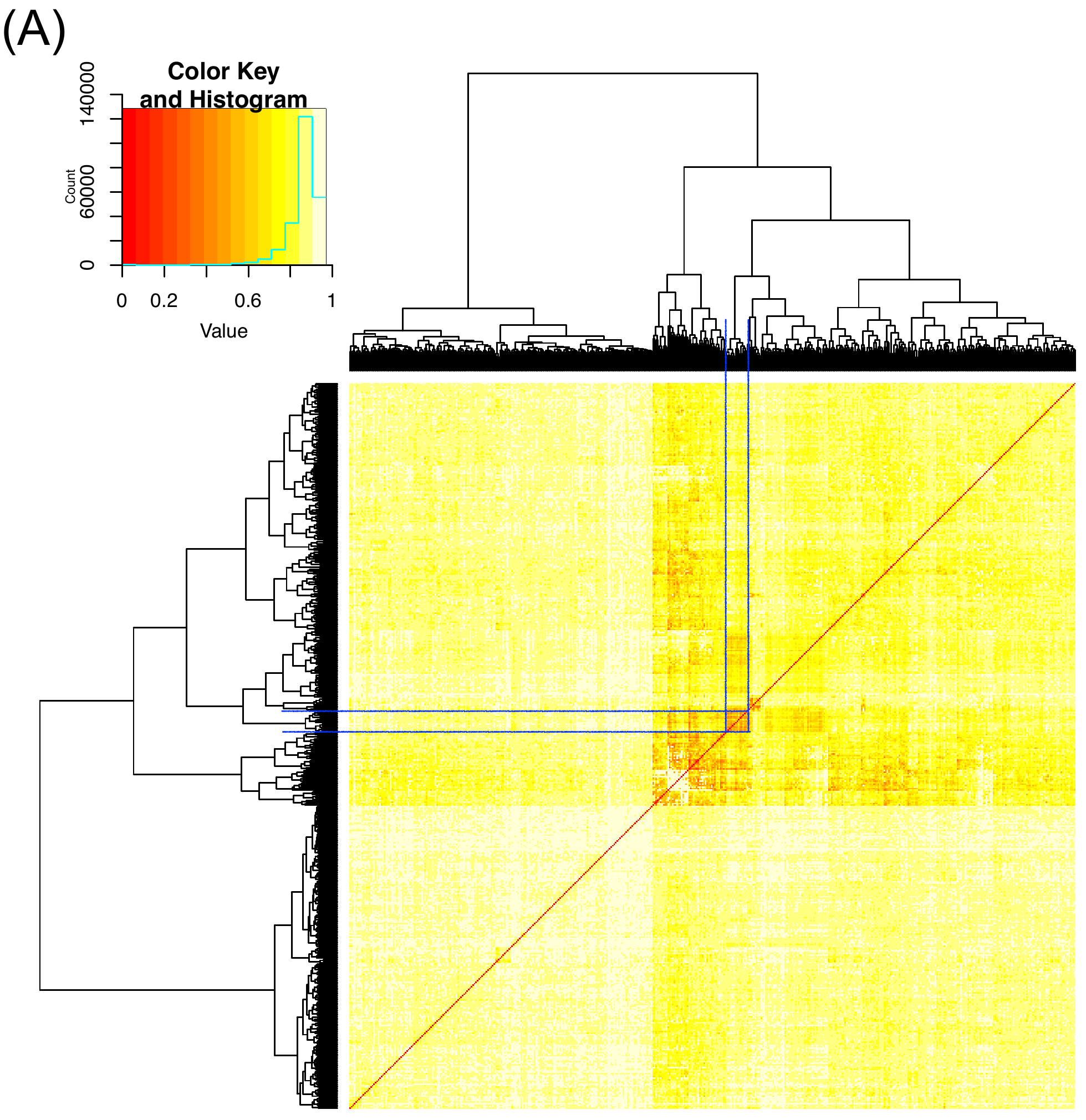}
  \end{tabular}
  \vspace{\floatsep}
  \vspace{\floatsep}
  \begin{tabular}{@{}c@{}}
    \includegraphics[width=.47\linewidth]{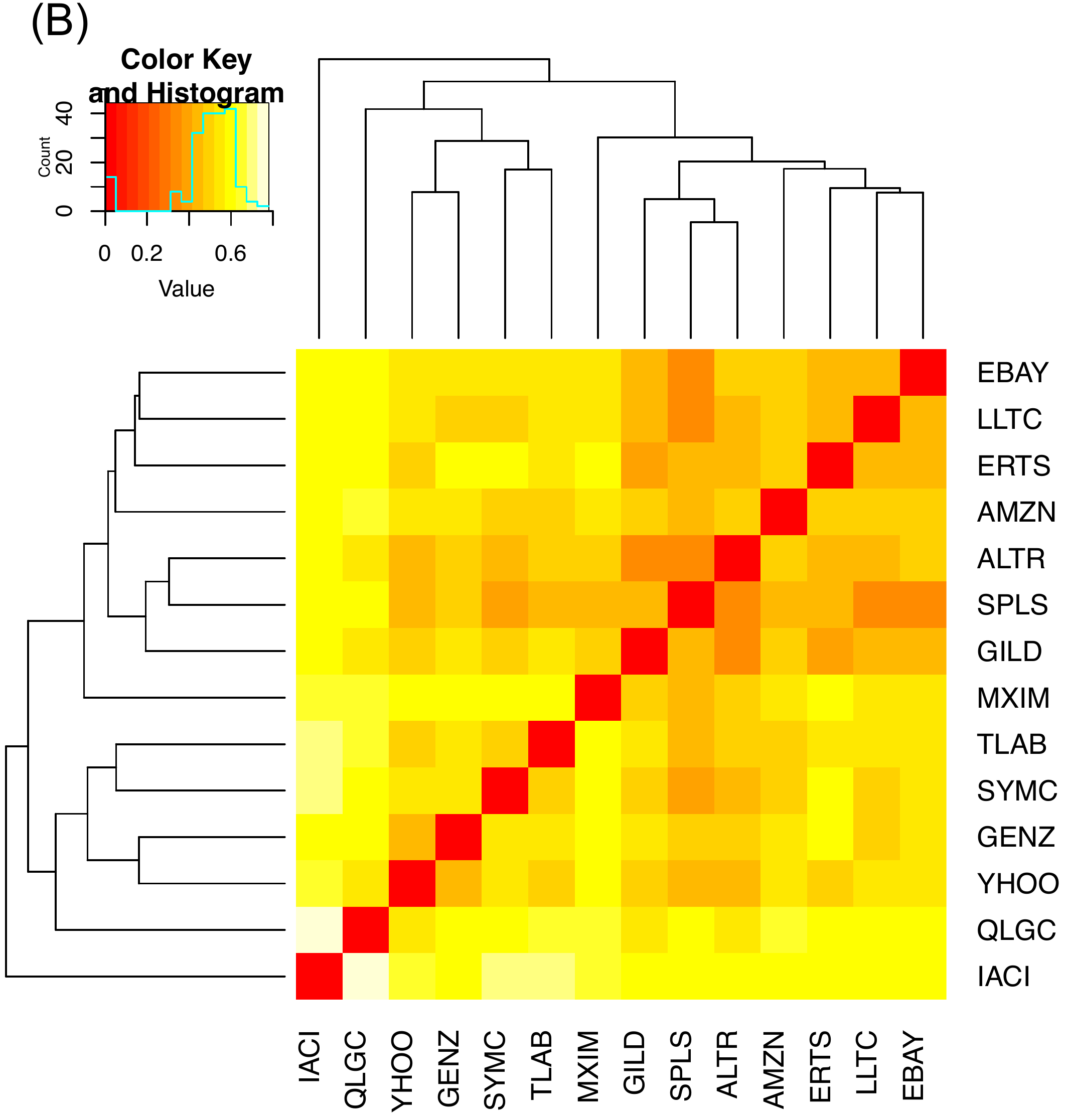}
\end{tabular}
\caption{Heatmap of the symmetric dissimilarity matrix with a hierarchical clustering tree imposed on the row and column sides; (A) a matrix for S$\&$P500; (B) a submatrix extracted from (A).}
\label{fig.heatmap}
\end{figure}

Another way to visualize the network is to transform the asymmetric matrix into a symmetric dissimilarity measure. The similarity between the $i$-th and the $j$-th nodes can be defined by the average of asymmetric TE values,
\begin{equation}
Sim(i,j)=(TE_{i\rightarrow j} + TE_{i\rightarrow j})/2
\end{equation}
If the range of similarity is rescaled between 0 and 1, the dissimilarity can be simply defined by,
\begin{equation}
Dis(i,j)=1-\frac{Sim(i,j)-\min_{i,j} Sim(i,j)}{\max_{i,j} Sim(i,j) - \min_{i,j} Sim(i,j)}
\end{equation}
So, the range of dissimilarity is scaled between 0 and 1. The symmetric dissimilarity matrix of S$\&$P500 is present in Figure\ref{fig.heatmap}(A) with a hierarchical clustering tree imposed on the row and column sides. The idea is similar to Multidimensional Scaling, which has been widely used to visualize the financial connectivity in a low-dimensional space \cite{33}. We claim that the dendrogram provided by hierarchical clustering is more informative in illustrating how the S$\&$P500 are hierarchically agglomerated from the bottom to the top according to their dissimilarity. Intuitively, companies under a similar industrial category should be merged into a small cluster branch. One of the branches with relatively low mutual distance is extracted and shown in Figure\ref{fig.heatmap}(B). It looks that the cluster mainly includes technology companies including internet retail (EBAY and AMZN), manufacturer of integrated circuits(LLTC), video games(ERTS), information technology(ALTR), network technology(TLAB), biotechnology(GILD and GENZ), etc. Besides, it is noticed that energy corporations, such as VLO, COP, and CHK, are also merged into a small cluster.

\begin{figure}[h]
    \centering
    \begin{subfigure}[b]{0.480\textwidth}
        \includegraphics[width=\textwidth]{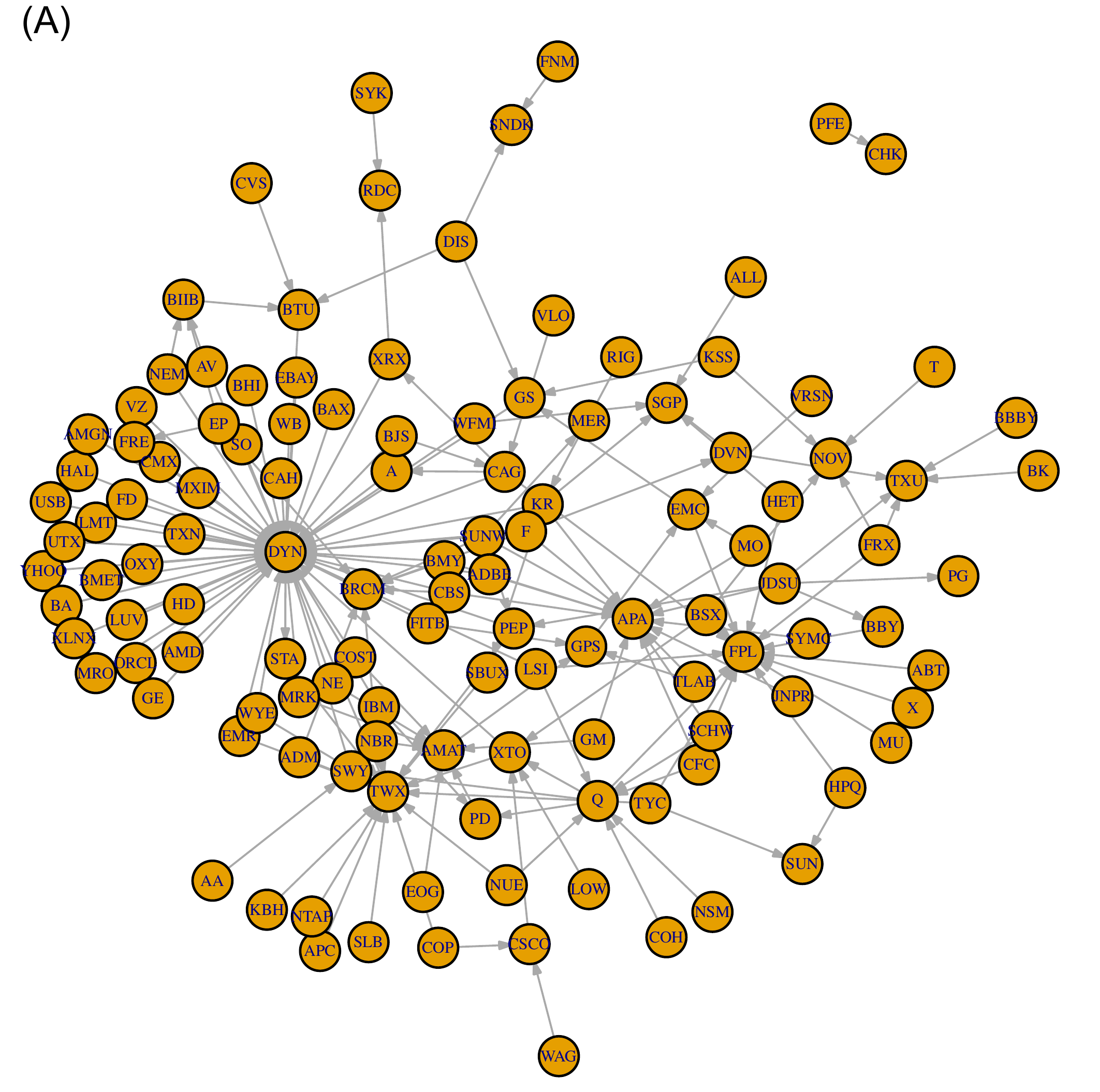}
    \end{subfigure}
    \quad 
    \begin{subfigure}[b]{0.480\textwidth}
        \includegraphics[width=\textwidth]{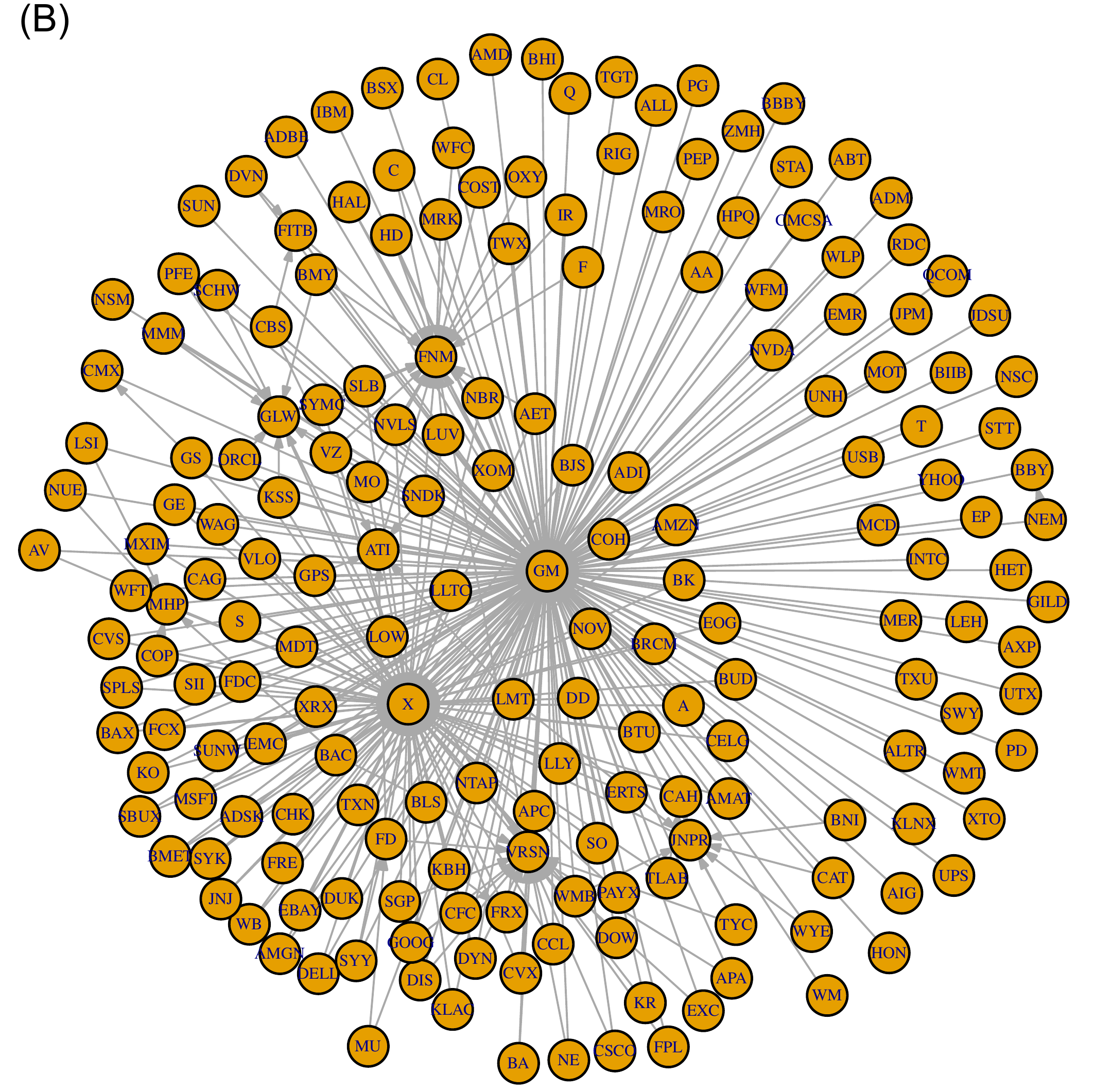}
    \end{subfigure}
\caption{Directed thresholding network of S\&P500; edge weight is measured based on Transfer Entropy with lag-5. (A) simple binning with 5-quantile cutting; (B) simple binning with 7-quantile cutting.}
\label{fig.simple_binning}
\end{figure}

As a comparison to the network constructed based on the decoding procedure, Transfer Entropy is calculated via simple binning with different cutting strategies, so a binning-based network is obtained accordingly. It shows that the resultant network is sensitive to the number of bins as the simple binning tends to overfit the error term in the high-frequent data (see Figure\ref{fig.simple_binning}) and the dissimilarity matrix shows no significant clustering structure.
However, the proposed decoding is able to model the volatility dynamic, and its pattern or number of code-states is stable to the price fluctuation.

\section{Conclusion}
Starting from a definition of large or relative extreme returns, we firstly propose a searching algorithm to segment stock returns into multiple levels of volatility phases. Then, we advocate a data-driven method, named encoding-and-decoding, to discover the embedded number of hidden states and represent the stock dynamics. By encoding the continuous observations into a sequence of 0-1 variables, a maximum likelihood approach is applied to fit the limiting distribution of the recurrence time series. Though the assumption of exchangeability within each hidden state is required, our numerical experiments show that the proposed approach still works when the assumption is slightly violated, for example, a weak transaction probability is imposed under the Markovian condition. This demonstration of robustness with respect to various conditions makes the proposed approach valuable in real-world finance researches and practices.

In real data application, it was reported by \cite{9} that stock returns are only exchangeable in a short period. With this assumption holds, our proposed method is implemented on high-frequent data to alleviate the serial dependency. Moreover, it is beneficial to investigate the fine-scale volatility, so the established network can illustrate which stocks stimulate or even promote volatility on others. It is also noted that the non-parametric regime-switching framework can work in conjunction with other financial models. For example, the forecasting procedure of HMM can be applied and improved with the help of encoding-and-decoding. Some future work like Peak Over Threshold(PoT) \cite{34} can be implemented to analyze the extreme value distribution based on the homogeneous regimes discovered by the proposed method.



\newpage

\section*{Appendix}
\appendix

\section{Simulation Appendix}\label{appendix:simulation}

\begin{figure}[h]
\centering
\includegraphics[width=5.2in]{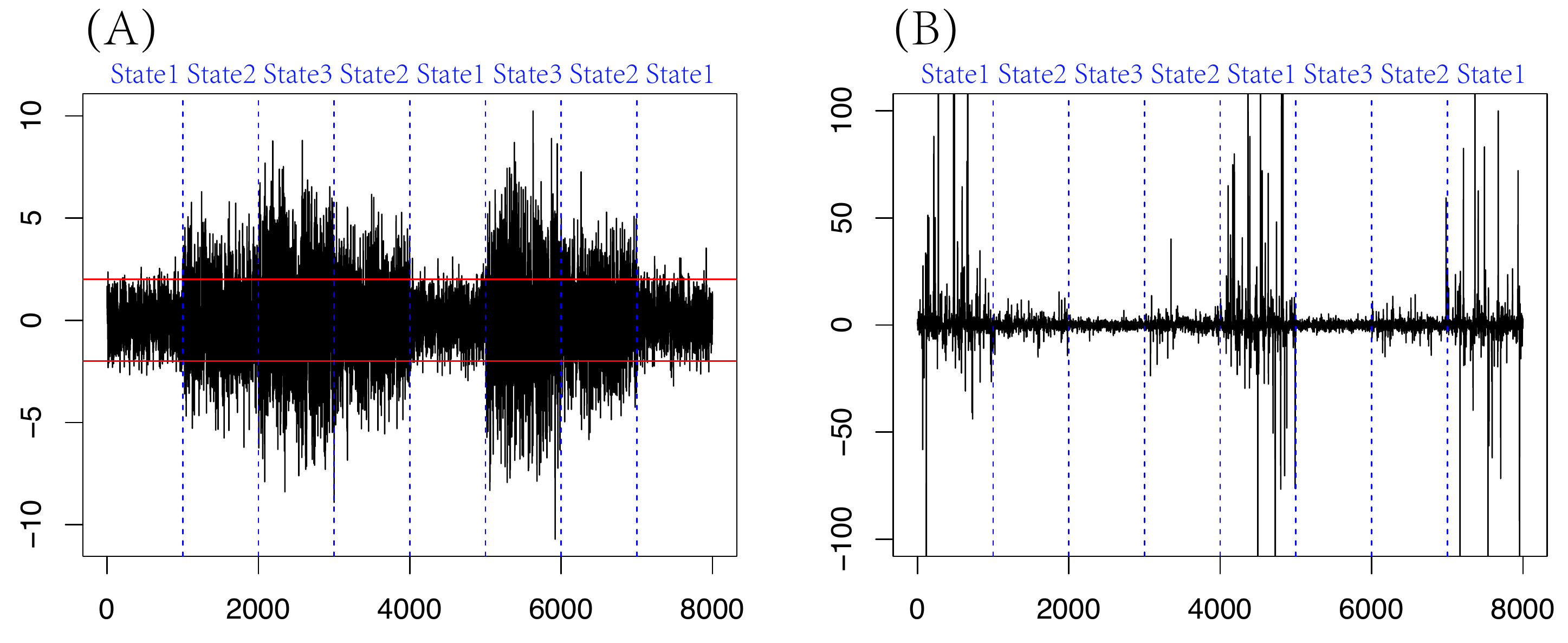}
\caption{Data is simulated via conditional distribution given a hidden state: (A) Gaussian distribution (B) student-t distribution. 8 underline phases alternates over time where 3 kinds of hidden states are embedded. The horizontal lines in (A) indicate the thresholds $l$ and $u$.}
\label{fig.simulation}
\end{figure}

Data is simulated under a regime-switching model with 3 hidden states embedded behind. Suppose that the observations (log returns) are $\{Y(t)\}_{t=1}^{8000}$ and there are 8 alternating segments over time:
\[S(t)= \begin{cases} 1 & \quad t \in [1,1000], [4001,5000], [7001,8000]\\
2 & \quad t \in [1001,2000], [3001,4000], [6001,7000]\\
3 & \quad t \in [2001,3000], [5001,6000]
\end{cases}\]
The index of the hidden states is alternating like $1,2,3,2,1,3,2,1$. In the first example, observations are generated of Gaussian distribution with mean 0 and variance varying under different states in Figure\ref{fig.simulation}(B), so
\[Y(t) \sim \begin{cases} N(0,\sigma_1^2) & \quad S(t)=1\\
N(0,\sigma_2^2) & \quad S(t)=2\\
N(0,\sigma_3^2) & \quad S(t)=3
\end{cases}\]
where standard deviations $\sigma_1=1$,$\sigma_2=2$, and $\sigma_3=3$. In the second example, heavy-tail distribution, student-t, is considered. The simulation is shown in Figure\ref{fig.simulation}(D) by
\[Y(t) \sim \begin{cases} t(df_1) & \quad S(t)=1\\
t(df_2) & \quad S(t)=2\\
t(df_3) & \quad S(t)=3
\end{cases}\]
where degree of freedoms $df_1=1$, $df_2=2$, and $df_3=5$. 

\section{Illustration of Alg.1}\label{appendix:alg.1}

In the Gaussian setting, we set $|u|=|l|=2$ which corresponds to 0.9 quantile of the observations; In the student-t setting, a larger threshold is considered, $|u|=|l|=3$, which corresponds to 0.95 quantile of the observations.
The recovered segment (marked in different colors) shows that with appropriate choice of thresholds, the proposed method can successfully detect the alternating hidden states. The error only appears around the change points. Besides, we obtain good estimations of emission probability under different hidden states. The estimators are $(0.0463, 0.3210, 0.4739)$ in the Gaussian setting and $(0.0420, 0.1008, 0.1997)$ in the student-t setting, respectively. They are close to the theoretical parameter
$$2*(\Psi_1(-2),\Psi_2(-2),\Psi_3(-2))=(0.0455, 0.3173, 0.5049)$$
and
$$2*(F_{t1}(-3),F_{t2}(-3),F_{t3}(-3))=(0.0300, 0.0954, 0.2048)$$
where $\Psi_i$ is the CDF of Gaussian distribution under the $i$-th hidden state, and $F_{ti}$ is the CDF of student-t distribution under the $i$-th hidden state, for $i=1,2,3$.

\begin{figure}[!h]
\centering
\begin{tabular}{@{}c@{}}
    \includegraphics[width=5.in]{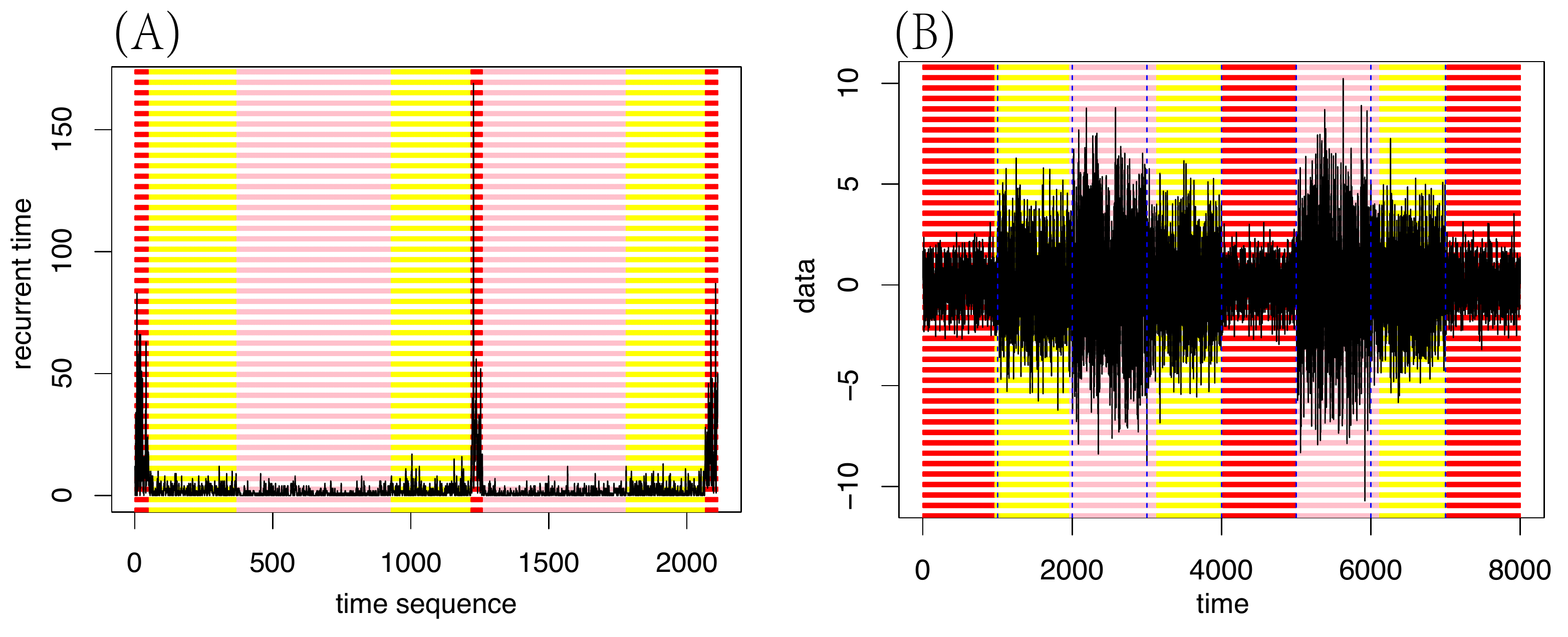}
  \end{tabular}
  \vspace{\floatsep}
  \begin{tabular}{@{}c@{}}
    \includegraphics[width=5.in]{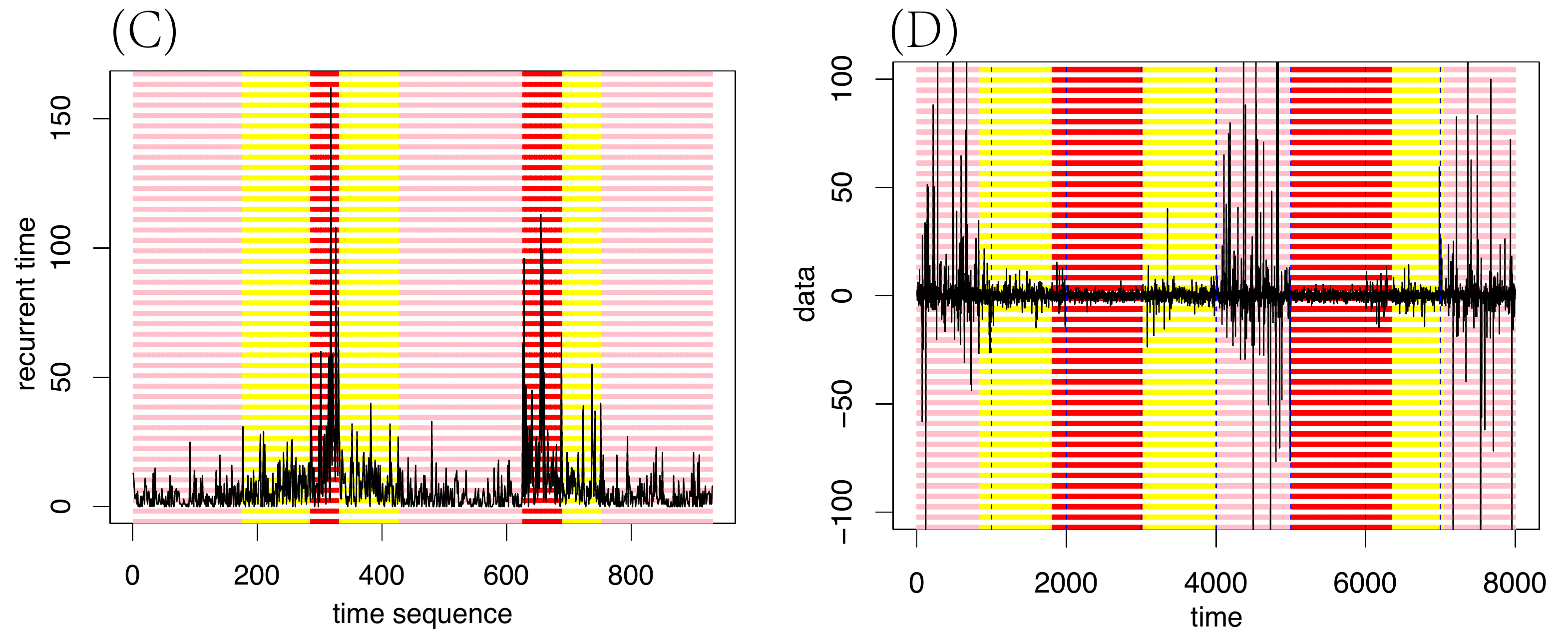}
\end{tabular}
\caption{Simulation with Normal distribution(A)(B), student-t distribution (C)(D). (A),(C): Recursive time; (B),(D): raw data with colored decoding states. ``red'', ``yellow'', and ``pink'' 3 colors indicates 3 different kinds of states.}
\label{fig.simulation_decoding}
\end{figure}

\break

\section{Calculation of Information Flows}\label{appendix:TE}

The proposed procedure is implemented to detect the volatility trajectory of the tick-by-tick returns. However, since the return patterns are recorded in transactions, the decoded sequences are not directly conjunct with each other. It is required to transform the decoded trajectories back into the real-time scale before calculating the Transfer Entropy. Suppose that the intensity level of volatility is indicated by ordinal number 1, 2, or 3 meaning low-, median-, or high-volatility state, respectively. If there exists a tiny time scale in which at most one transaction happens, a time unit can be labeled by symbol 0 for no transaction or an ordinal number if a transaction is present. Thus, the decoded pattern from different stocks can share the same chronological time. It seems that we attempt to choose a time scale as small as possible, but the pairwise dependency is weakened due to the increasing number of 0's. To balance the proportion of symbols and alleviate the sparsity, we summarize the recovered pattern by scanning a time block from the beginning to the end of the time axis to select the maximal ordinal number. So, uninformative 0's are filtered out, while volatility stages are kept. Suppose that the recovered symbol sequence is $\{S(t)\}_{t=1}^n$ where $S(t) \in \{0,1,2,3\}$. The sequence is then summarized via a time block with length $w$ by
\begin{equation}
S^*(t)=\max \{S(t) : t \in (t-\floor{\frac{w}{2}},t+\floor{\frac{w}{2}})\}
\end{equation}
for $t=\floor{\frac{w}{2}},...,n-\floor{\frac{w}{2}}$ where $\{S^*(t)\}_{t}$ is the summarized symbolic sequence. It is noted that a minute-level scale $w$ is too rough to reflect the tick-by-tick volatility pattern. A block with $w=5$-seconds is an admissible choice.

According to the way we summarize the symbolic trajectory, a non-linear measure is developed as a variant of TE. Different from the classic TE, this measure takes both lag and lead effects into account instead of only the lag effect. The lag-and-lead information flow from $X$ to $Y$ is defined by
\begin{equation}
TE_{X\rightarrow Y}^*(w)=\sum P(S^*_Y(t)=3,S^*_X(t)) log \frac{P(S^*_Y(t)=3|S^*_X(t))}{P(S^*_Y(t)=3)}
\end{equation}
We use $TE_{X\rightarrow Y}^*$ to differentiate it from the classic TE and $w$ is omitted without confusion. The measure is interpreted by how much uncertainty $Y$ is affected due to the lag-and-lead effect of $X$ such that $Y$ is under its volatility states (state3). The higher the value, the stronger the impact that $X$ promotes volatility phases of $Y$.

\break

\section{Graph and Table Appendix}\label{appendix:graph}

\begin{figure}[h]
\centering
\begin{tabular}{@{}c@{}}
    \includegraphics[width=4in]{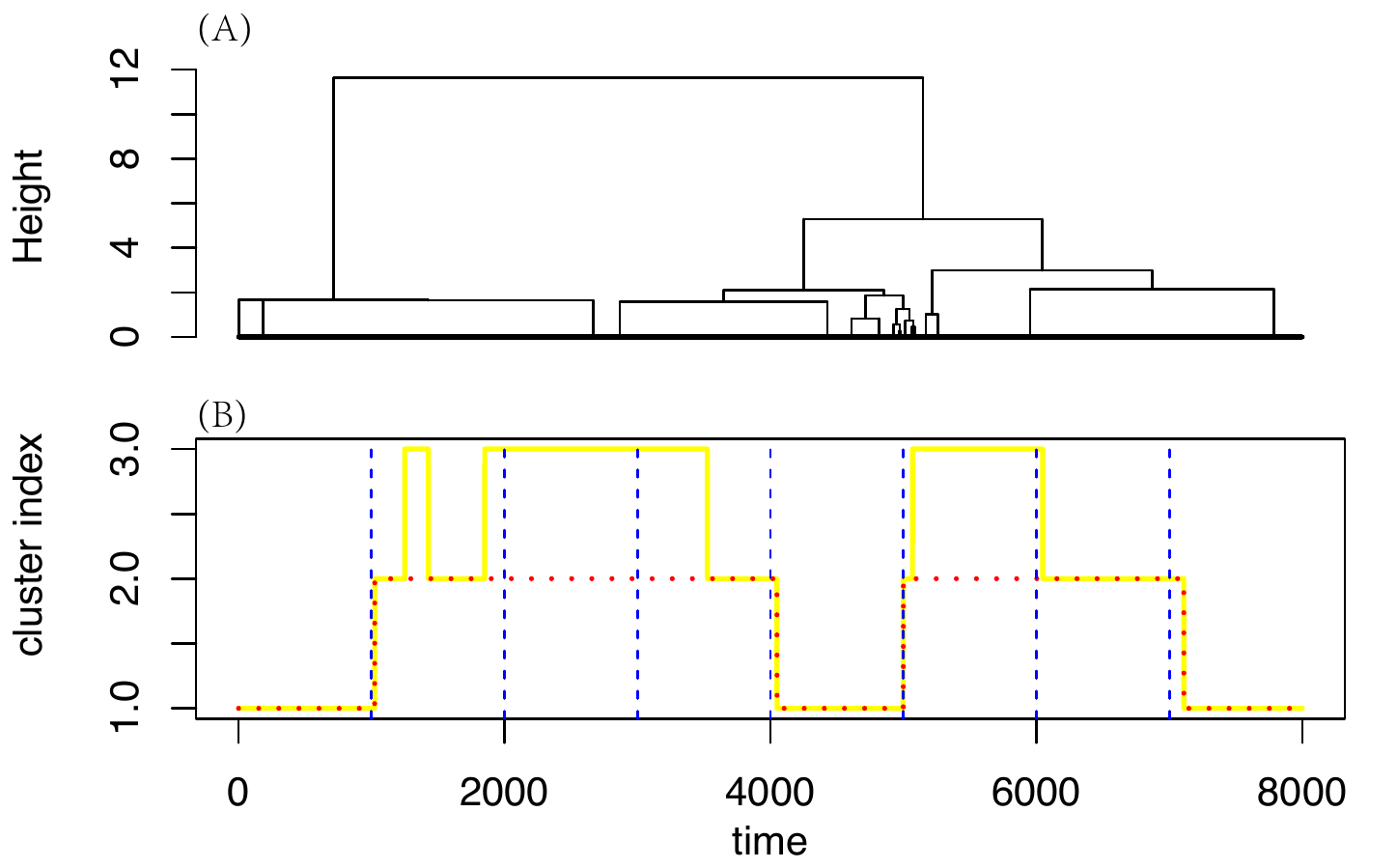}
  \end{tabular}
  \vspace{\floatsep}
  \begin{tabular}{@{}c@{}}
    \includegraphics[width=4in]{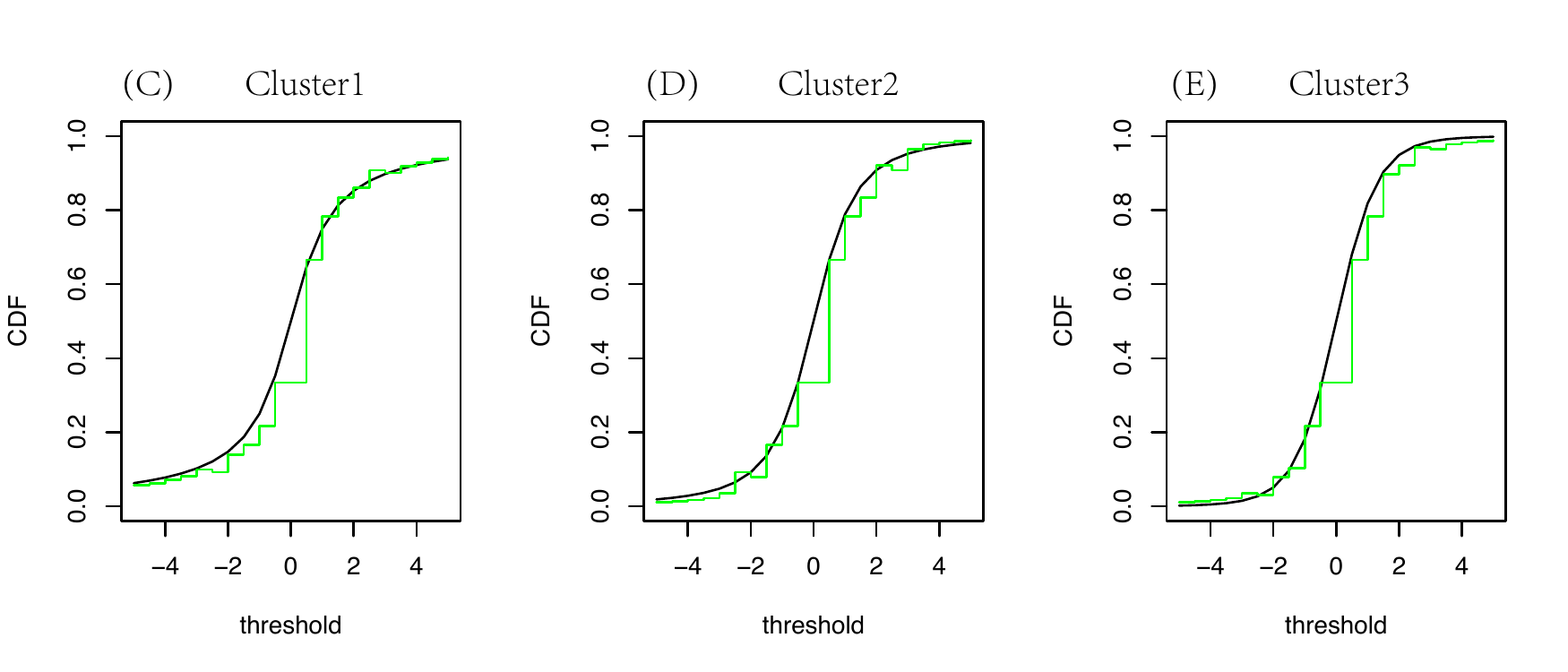}
\end{tabular}
\caption{2-states decoding results for simulated data in Appendix\ref{appendix:simulation}. (A) Hierarchical Clustering Tree; (B) cluster index switching over time; (C),(D),(E): estimated CDF versus true CDF, in cluster 1,2,3, respectively.}
\label{fig.simulation_2statesclustering}
\end{figure}

\begin{figure}[h]
\centering
\begin{tabular}{@{}c@{}}
    \includegraphics[width=4in]{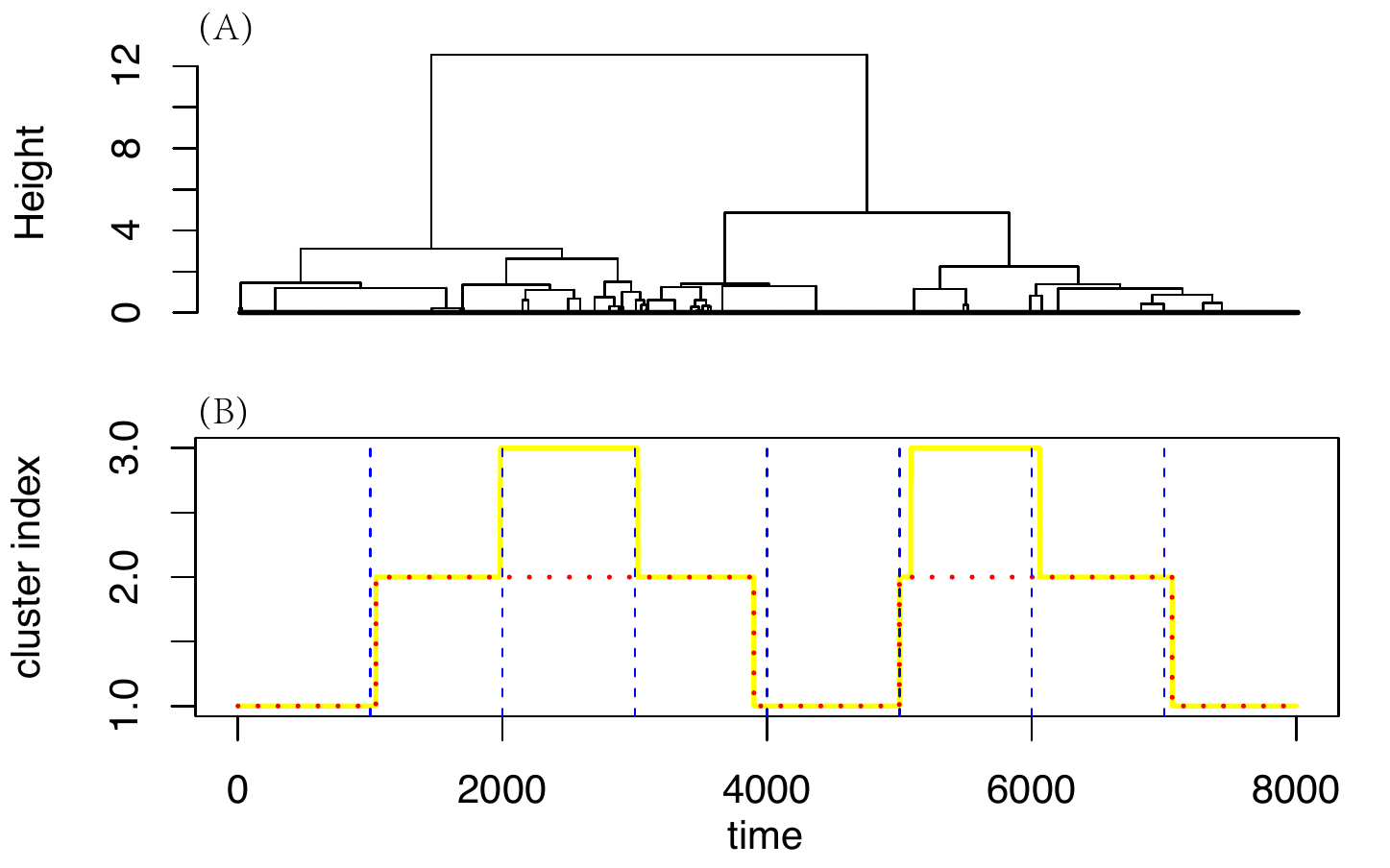}
  \end{tabular}
  \vspace{\floatsep}
  \begin{tabular}{@{}c@{}}
    \includegraphics[width=4in]{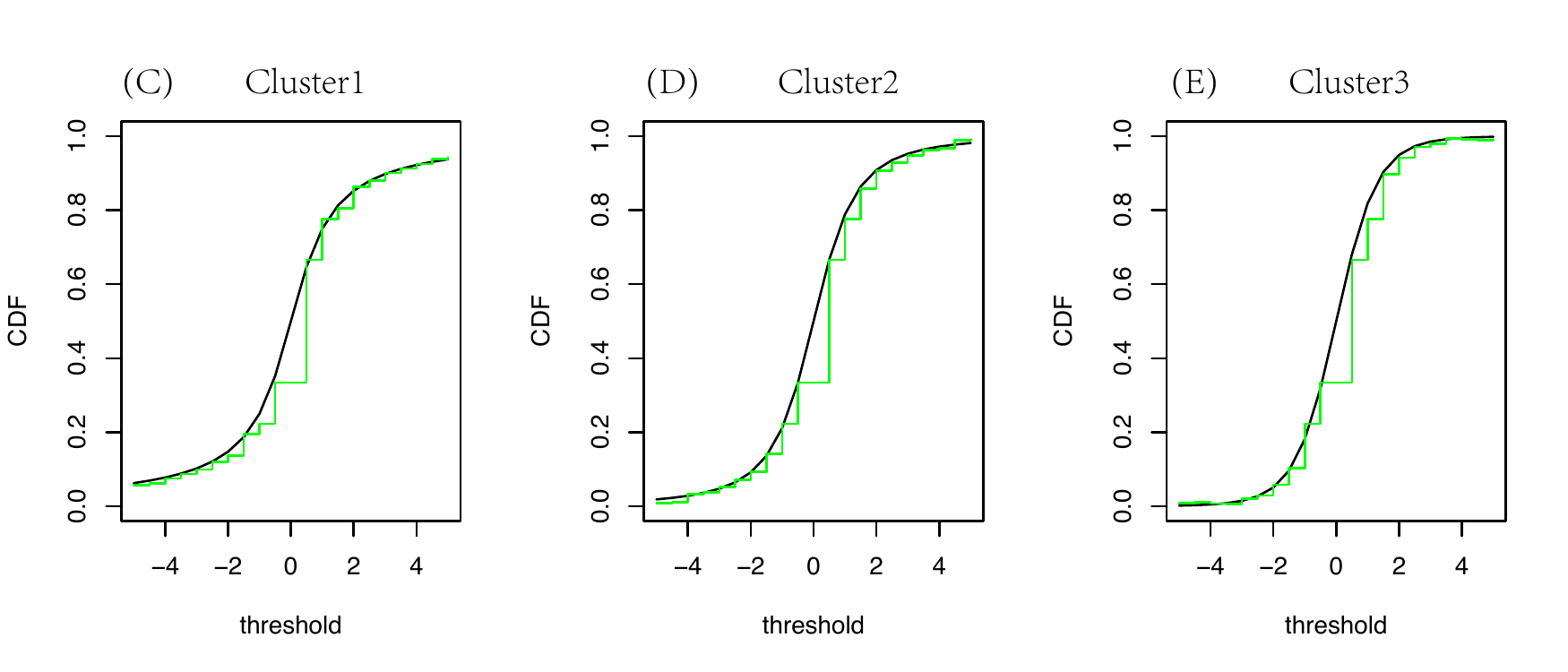}
\end{tabular}
\caption{4-states decoding results for simulated data in Appendix\ref{appendix:simulation}. (A) Hierarchical Clustering Tree; (B) cluster index switching over time; (C),(D),(E): estimated CDF versus true CDF, in cluster 1,2,3, respectively.}
\label{fig.simulation_4statesclustering}
\end{figure}

\begin{figure}[!h]
\centering
\includegraphics[width=3.5in]{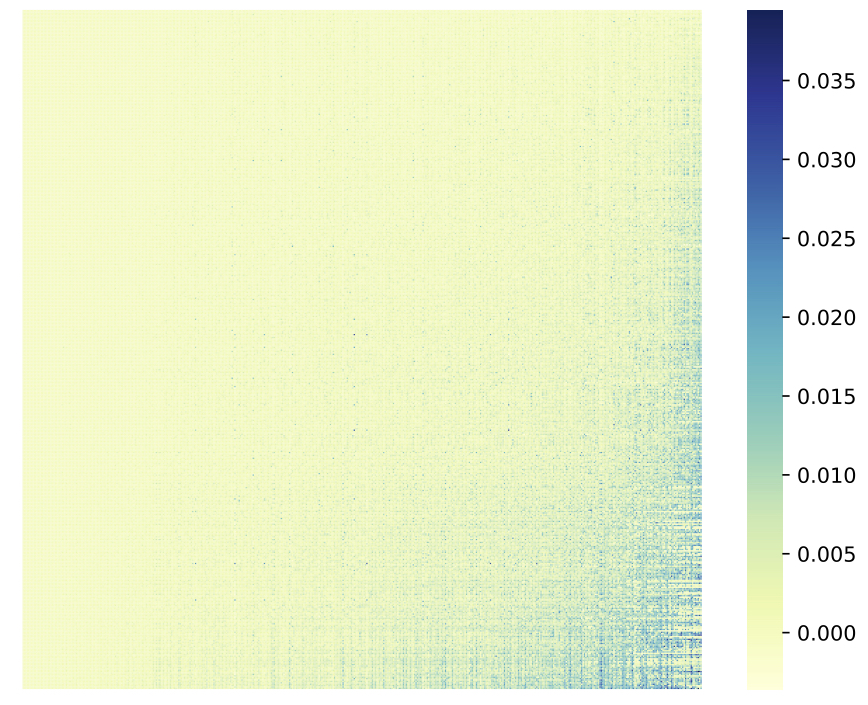}
\caption{Transfer Entropy matrix for S$\&$P500 in 2006. The rows and columns are rearanged such that the row sum and column sum are in ascending order.}
\label{fig.TEmatrix}
\end{figure}


\begin{table*}[h]
\centering
\caption{Top30 stocks with the strongest node strength}
\label{tab.top30}
\begin{tabular}{@{}lcclc@{}}
\hline
\multicolumn{2}{c}{incoming} & & \multicolumn{2}{c}{outgoing}\\
\cline{1-2} \cline{4-5}
Index & NS && Index & NS \\
\hline
EMC  & 4.3349 && TWX  & 3.3975  \\
BAC  & 4.2760 && BRCM & 3.3148  \\
NTAP & 4.2245 && NTAP & 3.2715  \\
JPM  & 4.0252 && GILD & 3.0749  \\
WFC  & 3.8934 && ALTR & 2.9504  \\
NBR  & 3.7955 && VLO  & 2.8755  \\
HON  & 3.6844 && EBAY & 2.8178  \\
BRCM & 3.6363 && HD   & 2.7764  \\
AIG  & 3.6327 && WMT  & 2.7619  \\
KBH  & 3.6230 && NVLS & 2.7501  \\
CAT      &3.6097  && CHK  &2.7336 \\
WB       &3.5598  && AMD  &2.7331 \\
CTX      &3.5570  && MXIM &2.7227 \\
WAG      &3.4599  && YHOO &2.6252 \\
BJS      &3.4538  && JNJ  &2.5732 \\
WLP      &3.4503  && SCHW &2.5519 \\
LOW      &3.3636  && IBM  &2.5448 \\
SWY      &3.3279  && XLNX &2.5334 \\
AXP      &3.2317  && BIIB &2.5313 \\
NOV      &3.1355  && LLTC &2.5274 \\
BUD      &3.1273  && MU   &2.5269 \\
CHK      &3.1227  && NVDA &2.5108 \\
DOW      &3.0809  && BMET &2.4964 \\
KSS      &3.0608  && TXN  &2.4894 \\
VLO      &3.0252  && C    &2.4635 \\
TWX      &3.0220  && ADBE &2.4633 \\
MO       &3.0072  && CELG &2.4574 \\
DE       &2.9712  && TGT  &2.4286 \\
COP      &2.9598  && KLAC &2.3873 \\
TRUE     &2.9564  && ESRX &2.3860 \\
\hline
\end{tabular}
\end{table*}

\end{document}